\documentclass[prd,11pt,twoside,a4paper,tightenlines,aps,nofootinbib,preprintnumbers]{revtex4}
\usepackage[dvips]{graphicx}
\usepackage{epsfig}
\usepackage{graphicx}
\usepackage{psfig}
\usepackage{amsfonts}
\usepackage{amssymb}
\usepackage[english]{babel}
\usepackage{amsmath}
\usepackage{latexsym,epsf,axodraw,pstricks}
\usepackage{epsfig,boxedminipage,floatflt}
\usepackage{amscd}
\newcommand{\bc}{\begin{center}}
\newcommand{\ec}{\end{center}}
\newcommand{\be}{\begin{equation}}
\newcommand{\ee}{\end{equation}}
\newcommand{\bea}{\begin{eqnarray}}
\newcommand{\eea}{\end{eqnarray}}
\newcommand{\half}{\frac{1}{2}}

\newcommand{\ts}[1]{{\mbox{\scriptsize #1}}}
\newcommand{\tinys}[1]{{\mbox{\tiny #1}}}
\newcommand{\bfig}{\begin{figure}[h]}
\newcommand{\efig}{\end{figure}}

\newcommand{\bi}{\begin{itemize}}
\newcommand{\ei}{\end{itemize}}
\newcommand{\ba}{\begin{align}}
\newcommand{\ea}{\end{align}}
\newcommand{\eref}[1]{(\ref{#1})}

\newcommand{\tpic}[1]{\;\parbox[c]{20pt}{\begin{picture}(20,30)(0,0)
\SetWidth{1.0}\SetScale{1.0} #1 \end{picture}}\;}
\newcommand{\pic}[1]{\;\parbox[c]{30pt}{\begin{picture}(30,30)(0,0)
\SetWidth{1.0}\SetScale{1.0} #1 \end{picture}}\;}

\newcommand{\picb}[1]{\;\parbox[c]{45pt}{\begin{picture}(45,30)(0,0)
\SetWidth{1.0}\SetScale{1.0} #1 \end{picture}}\;}
\newcommand{\picc}[1]{\;\parbox[c]{60pt}{\begin{picture}(60,30)(0,0)
\SetWidth{1.0}\SetScale{1.0} #1 \end{picture}}\;}

\newcommand{\Eye}{\parbox[c]{9pt}{\begin{picture}(9,5)(0,0)
\SetWidth{1.25}\SetScale{0.25}
\Oval(20,10)(9,9)(0)
\Vertex(11,10){3}
\Vertex(29,10){3}
\Line(1,10)(11,10)
\Line(29,10)(36,10)
\Line(11,10)(11,22)
\Line(29,10)(29,22)
\Line(27,24)(31,20)
\Line(31,24)(27,20)
\Line(9,24)(13,20)
\Line(13,24)(9,20)
\end{picture}}}

\newcommand{\Sigmaeye}{\Sigma_{\Eye}}

\begin{document}
\title{
Equilibration in 
$\varphi^4$ theory in 3+1 dimensions 
}
\author{Alejandro Arrizabalaga${}^{a,b}$, Jan Smit${}^a$, Anders
Tranberg${}^{a,c}$\\[-2ex] \ }

\affiliation{\mbox{\it ${}^a$ Institute for Theoretical Physics, University of
Amsterdam,} \\
\mbox{\it Valckenierstraat 65, 1018 XE Amsterdam, The Netherlands}\\ 
\mbox{\it ${}^b$ National Institute for Nuclear and High-Energy Physics
(NIKHEF),}\\ \mbox{\it Kruislaan 409, 1098 SJ, Amsterdam, The Netherlands}\\ 
\mbox{\it ${}^c$ Department of Physics and Astronomy,
University of Sussex,}\\
\mbox{\it Falmer,
Brighton,
East Sussex BN1 9QH.
UK.
}}

\date{\today}
\preprint{ITFA-2005-13, NIKHEF/2005-003}

\begin{abstract}
The process of equilibration in $\varphi^4$ theory is investigated for a homogeneous system in 3+1 dimensions and a variety of
out-of-equilibrium initial conditions, both in the symmetric and broken phase,
by means of the 2PI effective action. Two $\Phi$-derivable approximations including scattering effects are used: the two-loop and the ``basketball'',
the latter corresponding to the truncation of the 2PI effective action at $\mathcal{O}(\lambda^2)$. The
approach to equilibrium, as well as the
kinetic and chemical equilibration is investigated.  
\end{abstract}

\maketitle


\section{Introduction}
The approach to equilibrium is an important aspect of non-equilibrium dynamics.
In the context of particle physics,
a large part of 
the interest 
 derives from results of heavy-ion
collision experiments 
with the RHIC at Brookhaven.
The hydrodynamic description of the experiments suggests that there is early thermalization
\cite{Heinz:2001xi}, 
but a
short thermalization time seems
to contradict
traditional perturbative estimates \cite{Baier:2000sb,Molnar:2001ux}. 
This puzzle has been analyzed in terms of 
prethermalization \cite{Berges:2004ce}, and led to further
study of the
microscopic dynamical processes responsible for the equilibration of the quark-gluon plasma
\cite{Romatschke:2003ms,Arnold:2004ih,Arnold:2004ti,Rebhan:2004ur}. Understanding the dynamical
processes leading to equilibration in theories with simpler interactions 
may also shed some light on this issue. We focus in
this paper on the case of scalar $\varphi^4$ theory.
\par
An adequate method to study out-of-equilibrium dynamics from first principles 
is the closed-time-path formalism \cite{Schwinger:1951ex,Bakshi:1963,Bakshi:1963v2,Keldysh:1964}. This
scheme leads to causal equations of motion for the various correlation functions
that describe their time evolution. The initial conditions are
specified by a density matrix, which can be far from equilibrium. Most
applications of this method have focused on the study of the equations of
motion for the 1- and 2-point functions, known as the Kadanoff-Baym equations
\cite{KadanoffBaym}.
Close enough to equilibrium, where kinetic theory is applicable, the
Kadanoff-Baym equations have been used extensively, mostly in connection with
the study of transport phenomena and the derivation of effective Boltzmann
equations (see, for instance
\cite{Danielewicz:1984kk,Chou:1985,Calzetta:1988cq,Mrowczynski:1990bu,Greiner:1998vd,Blaizot:2001nr}
and references therein). 
\par
Far from equilibrium, kinetic theory is no longer valid, and simple
perturbation theory approaches fail to work due to the appearance of secular
terms (see for instance \cite{Cooper:1996bn}) and/or pinch singularities \cite{Altherr:1994fx}. These problems are
usually absent if one makes use of a self-consistent method, such as the Hartree approximation.
Unfortunately, real-time Hartree descriptions do not include sufficient scattering
between the field modes, and thus fail to describe the approach to
equilibrium. They are also not
``universal'', in the sense that the memory from the initial configuration is
not completely lost \cite{Aarts:2000wi}. An infinite number of conserved charges appear
that prevent the system from reaching a universal equilibrium state, independently of
the initial conditions. 
However, a Hartree ensemble approximation has been formulated to give an improved description of the early approach to equilibrium
\cite{Salle:2000hd,Salle:2002fu}.
\par
When the particle occupation numbers are large, another useful method far from equilibrium is the classical approximation. Interesting situations
where this occurs include 
preheating
 after cosmological inflation 
due to parametric resonance
\cite{Boyanovsky:1996sq,Khlebnikov:1997zt,Kofman:1997yn,Garcia-Bellido:1998wm}
or spinodal decomposition,
\cite{Boyanovsky:1995me,Boyanovsky:1996sq,Felder:2000hj,Felder:2001kt,Skullerud:2003ki,Arrizabalaga:2004iw}
 as well as the 
early stages of a heavy-ion collision
\cite{Krasnitz:1998ns,Krasnitz:1999wc,Krasnitz:2001qu,Lappi:2003bi}, where the gluon occupation numbers are as large as $\sim 1/\alpha_s$, up to a saturation scale
\cite{Mueller:1999wm,Baier:2000sb}. 
The classical approximation is not
good for describing quantum equilibration,
since the system does not move towards the quantum, but to the classical
 equilibrium state.
Nevertheless, the classical theory has
 been used to shed some light on the dynamics of equilibration and relaxation
 \cite{Aarts:1998kp,Aarts:1999zn,Aarts:2000mg,Boyanovsky:2003tc,Skullerud:2003ki}, as well as 
 a testground for comparison with various other approximation schemes
 \cite{Aarts:2000wi,Aarts:2001yn,Blagoev:2001ze,Arrizabalaga:2004iw}. 
 \par
A powerful scheme that takes into account both scattering and quantum effects is the two-particle
irreducible (2PI)
effective action \cite{Cornwall:1974vz,Calzetta:1988cq,Berges:2000ur}. The 2PI effective action furnishes a complete
representation of the theory in terms of the dressed 1- and 2- point functions. 
The exact equations of motion describing the time evolution of these correlation functions are obtained by a variational principle on the 2PI
effective action functional. Various approximations to the equations of motion can be obtained if one applies the variational method to a truncated
version of the action. By construction, these are self-consistent and thus
free of 
secular problems. The approximation can be improved, in principle, by 
truncating the 2PI effective action at higher order in some expansion
parameter.
\par The main advantages of the 2PI effective action approach stem from the fact that the approximations are performed on the level of a
functional. For that reason the approximations have also been called Functional-derivable, or $\Phi$-derivable. The 2PI effective action functional (and any truncation thereof) is, by construction, invariant under global transformations of the 1- and 2-point functions. The
variational procedure on any truncation guarantees that the global symmetries are still preserved
by the equations of motion. Their associated Noether currents are thus conserved. In particular, this implies
that the derived equations of motion conserve energy, as well as global
charges \cite{Baymconserve,Ivanov:1998nv,Bedingham:2003jt}. This
is a very important feature when studying out-of-equilibrium processes, where
most other quantities evolve in complicated ways. The $\Phi$-derivable approximations to the 2PI effective action constitute
thus a very convenient method for studying equilibration. 
\par
In
recent years, approximations based on the 2PI effective action have been
applied succesfully to the study of non-equilibrium real-time dynamics. In
the context of scalar theories, studies of equilibration have been carried
out in the 3-loop $\Phi$-derivable approximation for $\varphi^4$ theory, in
the symmetric phase, both in 1+1 \cite{Berges:2000ur,Aarts:2001qa} and 2+1 dimensions
\cite{Juchem:2003bi}. In the broken phase, the case of 1+1 dimensions has
also been discussed in \cite{Cooper:2002ze,Cooper:2002qd}. Similar
studies of thermalization have been performed in the $O(N)$ model in 1+1 dimensions, both at
next-to-leading (NLO) order 
in a $1/N$ expansion \cite{Berges:2001fi,Aarts:2002dj}, and in the bare vertex
approximation \cite{Cooper:2002ze,Cooper:2002qd,Mihaila:2003mh}.  All these
analyses, which include scattering, show that the system indeed equilibrates, with the equilibrium state independent of
the initial conditions.
Comparing with the loop expansion, the $1/N$ expansion has the advantage that it is applicable in situations where
large particle numbers are generated. This has allowed the study of
interesting phenomena, such as parametric resonance
\cite{Berges:2002cz} or spinodal decomposition during a phase transition \cite{Arrizabalaga:2004iw}. The
studies in \cite{Berges:2002cz} and \cite{Arrizabalaga:2004iw} were done for the
$O(N)$ model at NLO, in 3+1 dimensions.
Methods based on the 2PI effective action have also been applied to theories
with fermions, in 3+1 dimensions \cite{Berges:2002wr,Berges:2004ce}. The extension
of 
the 2PI effective action methods for gauge theories, however, is not
straightforward due to a residual dependence on the choice of gauge condition
\cite{Arrizabalaga:2002hn,Carrington:2003ut,Calzetta:2004sh,Andersen:2004re}. 
\par
In this paper, we use the loop-expansion of the 2PI effective action to study
the approach to equilibrium for the case of a real scalar $\varphi^4$ theory in 3+1
dimensions,
both
in the symmetric and broken phase, complementing in this manner the investigations in
\cite{Berges:2000ur,Cooper:2002qd,Juchem:2003bi}. 


\section{2PI loop expansion of $\varphi^4$ theory}
\begin{figure}[h]
\epsfig{file=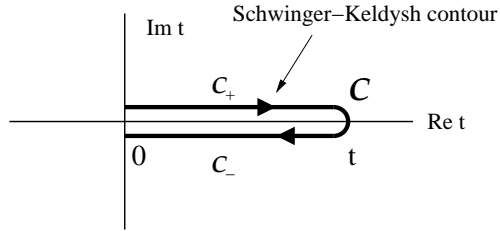, width=.40\textwidth,clip}
\caption{Schwinger-Keldysh contour.}
\label{fig:ctp}
\end{figure}

For $\varphi^4$ theory we write the action as
\begin{equation}
S[\varphi]=\int_{\mathcal{C}}d^4x\,\left[\half
 \partial_\mu\varphi(x)\partial^\mu\varphi(x)-\half
m^2 \varphi(x)^2-\frac{\lambda}{4!}\varphi(x)^4\right],
\label{cubiclagrangian2}
\end{equation}
The subscript $\mathcal{C}$ indicates that the integrations are performed 
 along
the real-time Schwinger-Keldysh contour, running from an initial time $t_0$ to
time $t$ along $\mathcal{C}_+$ and going back to $t_0$ along $\mathcal{C}_-$
(see 
 Fig. 
\ref{fig:ctp}). The formulation of the theory along the real-time
contour $\mathcal{C}$ is appropiate for studying non-equilibrium problems \cite{Schwinger:1961,Keldysh:1964,Chou:1985}. 
\par
The system can be in two distinct phases: the \emph{symmetric phase} (the vacuum field expectation value $v$ is
$v=0$) which occurs for $m^2 > 0$, and the \emph{broken phase} ($ v\neq
0$) for $m^2 < 0$. At tree level, the vacuum expectation value in the broken
phase is given by $v_\ts{tree}=\sqrt{6|m^{2}|/\lambda}$.  
\par 
The complete information about the theory can be written in terms of the
2PI effective action, which depends
explicitly on the full connected 1- and 2-point functions
$\phi(x)\equiv\langle \varphi(x) \rangle$ and $G(x,y)\equiv\langle
T_{\mathcal{C}} \varphi(x)\,\varphi(y)\rangle-\phi(x)\phi(y)$. For scalar
$\lambda \varphi^4$ theory, the 2PI
effective action functional can be written as \cite{Cornwall:1974vz}
\be
\Gamma[\phi,G]=S[\phi]-\frac{i}{2}\mbox{Tr}\ln
G+\frac{i}{2}\mbox{Tr}\Big[(G_0^{-1}-G^{-1})\cdot G\Big]+\Phi[\phi,G],
\label{2PIfunctional}
\ee
with
\be
iG_0^{-1}(x,y)=\frac{\delta^2 S[\phi]}{\delta \phi(x) \delta \phi(y)}=\left(
-\partial_x^2-m^2-\frac{\lambda}{2}\phi(x)^2\right)\delta_\mathcal{C}(x,y).
\ee
The contour delta function $\delta_{\mathcal{C}}(x,y)$ is given by  
\begin{equation}
\delta_\mathcal{C}(x,y)=
\begin{cases}
1 & \mbox{if $x=y$ and ${x,y} \in \mathcal{C}_+$,}\\
-1 & \mbox{if $x=y$ and ${x,y} \in \mathcal{C}_-$,}\\
0 & \mbox{otherwise.}
\end{cases}
\end{equation}
The functional $\Phi$ comprises the sum of the closed two-particle-irreducible (2PI) \emph{skeleton
diagrams}. Up to three loops it is given by
\begin{equation}
i\Phi[\phi,G]=
\frac{1}{8}
\tpic{
\Oval(15,7.5)(7.5,7.5)(0)
\Oval(15,22.5)(7.5,7.5)(0)
\Vertex(15,15){2}
}
+\frac{1}{12}
\picb{
\GCirc(22.5,15){11.5}{1}
\Line(2,15)(43,15)
\Vertex(11,15){2}
\Vertex(34,15){2}
\Line(0,17)(4,13)
\Line(0,13)(4,17)
\Line(45,17)(41,13)
\Line(45,13)(41,17)
}
+\frac{1}{48}
\pic{
\GCirc(15,15){15}{1}
\Oval(15,15)(6.5,15)(0)
\Vertex(0,15){2}
\Vertex(30,15){2}
}
+\frac{1}{24}\picc{
\GCirc(30,15){15}{1}
\CArc(15,30)(15,270,0)
\CArc(45,30)(15,180,270)
\Vertex(30,30){2}
\Vertex(15,15){2}
\Vertex(45,15){2}
\Line(2,15)(15,15)
\Line(45,15)(58,15)
\Line(0,17)(4,13)
\Line(4,17)(0,13)
\Line(60,17)(56,13)
\Line(56,17)(60,13)
}
+\frac{1}{24}\parbox[c]{40pt}{\begin{picture}(40,40)(0,0)
\SetWidth{1.0}\SetScale{1.0}
\GCirc(20,20){14}{1}
\Vertex(10,10){2}
\Vertex(30,10){2}
\Vertex(10,30){2}
\Vertex(30,30){2}
\Line(0,0)(40,40)
\Line(0,40)(18,22)
\Line(22,18)(40,0)
\Line(0,4)(4,0)
\Line(0,36)(4,40)
\Line(40,36)(36,40)
\Line(36,0)(40,4)
\end{picture}}.
\label{3looptruncation}
\end{equation}
The Feynman rules for these diagrams are given by
\begin{equation}
\tpic{
\Line(5,5)(25,25)
\Line(5,25)(25,5)
\Vertex(15,15){2}
}=-i\lambda, \qquad 
\pic{
\Line(0,15)(30,15)
\Text(0,8)[ct]{$\ts{$x$}$}
\Text(30,8)[ct]{$\ts{$y$}$}
}=G(x,y), \qquad 
\pic{
\Line(5,15)(23,15)
\Text(5,8)[ct]{$\ts{$x$}$}
\Line(25,17)(21,13)
\Line(21,17)(25,13)
}=\phi(x).
\end{equation}
In this manner, the functional $\Phi[\phi,G]$ is
\begin{multline}
\Phi[\phi,G]=-\frac{\lambda}{8}\int_{\mathcal{C}}d^4x\; G(x,x)^2
+i\frac{\lambda^2}{12}\int_{\mathcal{C}}d^4x\int_{\mathcal{C}}d^4y\;
\phi(x)G(x,y)^3 \phi(y)+i\frac{\lambda^2}{48}\int_{\mathcal{C}}d^4x\int_{\mathcal{C}}d^4y\; G(x,y)^4
+\ldots
\end{multline}
The 2PI effective action $\Gamma[\phi,G]$ provides an exact representation of the full theory. 
Considering only a finite number of terms in the series of diagrams in $\Phi$
leads to a truncated action, from which approximate ``physical'' 1- and
2-point functions can be obtained by a variational procedure. As a result of this, a resummation of effects from higher
orders in perturbation theory is performed.
\par
In this paper we shall investigate the truncations of the 2PI effective action up to three loops, in particular up to
${\mathcal{O}}(\lambda^2)$. The various truncations considered and their corresponding truncated functionals
$\Phi_\ts{tr}$ are displayed in table \ref{table:truncations}. 
The organization of the truncations discussed is based on the superficial counting of loops and/or vertices in the diagrams of $\Phi$, i.e. no assumption is taken on the coupling constant dependence of $\phi$ or $G$.
\begin{table}
\begin{tabular}{|c|c|c|}\hline
\emph{Truncation} & $\quad$ \emph{Order} $\quad$ & $i\Phi_\tinys{tr}[\phi,G]$ \\ \hline
Hartree approximation & ${\mathcal{O}}(\lambda)$ &
$\parbox[c]{1pt}{\begin{picture}(1,25)(0,0)
\SetWidth{1.0}\SetScale{1.0} 
\end{picture}}

\frac{1}{8}
\parbox[c]{20pt}{\begin{picture}(20,20)(0,0)
\SetWidth{1.0}\SetScale{1.0} 
\Oval(10,5)(5,4.5)(0)
\Oval(10,15)(5,4.5)(0)
\Vertex(10,10){1.5}
\end{picture}}
$ \\
Two-loop approximation & 2 loops & 
$
\parbox[c]{1pt}{\begin{picture}(1,25)(0,0)
\SetWidth{1.0}\SetScale{1.0} 
\end{picture}}

\frac{1}{8}
\parbox[c]{20pt}{\begin{picture}(20,20)(0,0)
\SetWidth{1.0}\SetScale{1.0} 
\Oval(10,5)(5,4.5)(0)
\Oval(10,15)(5,4.5)(0)
\Vertex(10,10){1.5}
\end{picture}}
+
\frac{1}{12}\;
\parbox[c]{30pt}{\begin{picture}(30,20)(0,0)
\SetWidth{1.0}\SetScale{1.0}
\GCirc(15,10){7}{1}
\Line(2,10)(28,10)
\Vertex(8,10){1.5}
\Vertex(22,10){1.5}
\Line(0.5,11.5)(3.5,8.5)
\Line(0.5,8.5)(3.5,11.5)
\Line(29.5,11.5)(26.5,8.5)
\Line(29.5,8.5)(26.5,11.5)
\end{picture}}
$ \\
``Basketball'' approximation & ${\mathcal{O}}(\lambda^2)$ & 
$
\parbox[c]{1pt}{\begin{picture}(1,25)(0,0)
\SetWidth{1.0}\SetScale{1.0} 
\end{picture}}

\frac{1}{8}
\parbox[c]{20pt}{\begin{picture}(20,20)(0,0)
\SetWidth{1.0}\SetScale{1.0} 
\Oval(10,5)(5,4.5)(0)
\Oval(10,15)(5,4.5)(0)
\Vertex(10,10){1.5}
\end{picture}}
+
\frac{1}{12}\;
\parbox[c]{30pt}{\begin{picture}(30,20)(0,0)
\SetWidth{1.0}\SetScale{1.0}
\GCirc(15,10){7}{1}
\Line(2,10)(28,10)
\Vertex(8,10){1.5}
\Vertex(22,10){1.5}
\Line(0.5,11.5)(3.5,8.5)
\Line(0.5,8.5)(3.5,11.5)
\Line(29.5,11.5)(26.5,8.5)
\Line(29.5,8.5)(26.5,11.5)
\end{picture}}
+\frac{1}{48}\;
\parbox[c]{30pt}{\begin{picture}(20,20)(0,0)
\SetWidth{1.0}\SetScale{1.0}
\GCirc(10,10){10}{1}
\Oval(10,10)(4,9.5)(0)
\Vertex(0,10){1.5}
\Vertex(20,10){1.5}
\end{picture}}
$ \\ \hline
\end{tabular}
\caption{Truncations of the 2PI effective action.}
\label{table:truncations}
\end{table}
\par In our analysis 
using the three-loop ``basketball approximation'' we have neglected the other
three-loop diagrams 
\begin{equation}
\picc{
\GCirc(30,15){15}{1}
\CArc(15,30)(15,270,0)
\CArc(45,30)(15,180,270)
\Vertex(30,30){2}
\Vertex(15,15){2}
\Vertex(45,15){2}
\Line(2,15)(15,15)
\Line(45,15)(58,15)
\Line(0,17)(4,13)
\Line(4,17)(0,13)
\Line(60,17)(56,13)
\Line(56,17)(60,13)
}\quad,\quad
\parbox[c]{40pt}{\begin{picture}(40,40)(0,0)
\SetWidth{1.0}\SetScale{1.0}
\GCirc(20,20){14}{1}
\Vertex(10,10){2}
\Vertex(30,10){2}
\Vertex(10,30){2}
\Vertex(30,30){2}
\Line(0,0)(40,40)
\Line(0,40)(18,22)
\Line(22,18)(40,0)
\Line(0,4)(4,0)
\Line(0,36)(4,40)
\Line(40,36)(36,40)
\Line(36,0)(40,4)
\end{picture}}.
\label{neglecteddiagrams}
\end{equation}
These diagrams are respectively of superficial order $\mathcal{O}(\lambda^3\phi^2)$ and $\mathcal{O}(\lambda^4 \phi^4)$. In the
symmetric phase, where $\phi\sim 0$, they can be safely
neglected. In the broken phase, however, $\phi\sim v
_\ts{tree}\sim
\mathcal{O}(\lambda^{-1/2})$ and thus both diagrams become
$\mathcal{O}(\lambda^2)$. In this situation it is not 
clear whether these contributions can
be ignored. Due to the difficulty in treating the above diagrams
numerically, we decided to neglect them in our analysis. Part of the first diagram in \eref{neglecteddiagrams}
can be
recovered 
at NLO 
in a $1/N$-expansion \cite{Aarts:2002dj}.


\section{Equations of motion}
In the formulation on the real-time contour ${\mathcal{C}}$, a
$\Phi$-derivable approximation to the 2PI effective action $\Gamma$ leads to
equations of motion for the 1- and 2-point functions. Indeed, solving the stationarity conditions
\begin{equation}
\frac{\delta \Gamma[\phi,G]}{\delta \phi}=0,\qquad \frac{\delta \Gamma[\phi,G]}{\delta G}=0,
\end{equation}
leads to the equation for the mean field
\be
\frac{\delta S[\phi]}{\delta \phi(x)}+\half\lambda G(x,x) \phi(x)=-\frac{\delta
\Phi[\phi,G]}{\delta \phi(x)},
\label{eomphi}
\ee
and for the 2-point function
\be
\delta_\mathcal{C}(x,y)=\int_{\mathcal{C}}d^4z\,G_0^{-1}(x,z)G(z,y)+i\int_{\mathcal{C}}d^4z\,\Sigma(x,z)G(z,y).
\label{eomG}
\ee
The self-energy $\Sigma(x,y)$ is given, in terms of the functional $\Phi$, by\footnote{Our convention for the
self-energy $\Sigma$ is that it appears, formally, as a positive contribution to the
mass. In particular, it is given in terms of the self-energy $\Sigma_B$ used in
\cite{Berges:2001fi} by $\Sigma=i\Sigma_B$.}
\be
\Sigma(x,y)=-2\frac{\delta \Phi[\phi,G]}{\delta G(y,x)}.
\ee
To the order considered here, the self-energy $\Sigma$ is determined from the truncated functional $\Phi^\ts{tr}[\phi,G]$. To $\mathcal{O}(\lambda^2)$ one finds
\begin{equation}
\Sigma[\phi,G]=i\Big[\,
\half\;
\parbox[c]{25pt}{
\begin{picture}(25,32.5)(0,0)
\SetWidth{1.0}\SetScale{1.0}
\Line(0,10)(25,10)
\Vertex(12.5,10){2}
\Oval(12.5,20)(7.5,10)(90)
\end{picture}}\;
+\half\;
\picb{
\GCirc(22.5,15){11.5}{1}
\Line(0,15)(11,15)
\Line(34,15)(45,15)
\Vertex(11,15){2}
\Vertex(34,15){2}
\Line(11,15)(10,27.5)
\Line(12,29.5)(8,25.5)
\Line(8,29.5)(12,25.5)
\Line(34,15)(35,27.5)
\Line(37,29.5)(33,25.5)
\Line(33,29.5)(37,25.5)
}\;
+\frac{1}{6}\;
\picb{
\GCirc(22.5,15){12.5}{1}
\Line(0,15)(45,15)
\Vertex(10,15){2}
\Vertex(35,15){2}
}\;
\Big]
\label{SEth}
\end{equation}
For the case of the Hartree approximation (see Table \ref{table:truncations}), only the first diagram in \eref{SEth} (the ``leaf'' 
diagram) enters in $\Sigma$. For the case of the 
two-loop 
and ``basketball'' approximations respectively, the second and third diagrams in \eref{SEth} (the ``eye'' and the ``sunset'') have to be taken into account. For the study of nonequilibrium dynamics these are important diagrams as they account for scattering 
and hence can
lead to equilibration.
\par
The self-energy can be split up into a local and a nonlocal part, 
\be
\Sigma(x,y)=\Sigma^{\rm l}(x) \delta_{{\mathcal{C}}}(x,y)+\Sigma^{\rm nl}(x,y),
\ee
with 
\begin{align}
\Sigma^{\rm l}(x)&=\frac{\lambda}{2}
G(x,x),\label{sigmaLth}\\
\Sigma^{\rm nl}(x,y)&=-i\frac{\lambda^2}{2}\phi(x)G(x,y)^2\phi(y)-i\frac{\lambda^2}{6} G(x,y)^3. \label{sigmaNL}
\end{align}
The quantities entering in the equations of motion \eref{eomphi} and
\eref{eomG} are defined in the real-time contour $\mathcal{C}$. For the
non-local quantities, such as $G(x,y)$ and $\Sigma^{\rm nl}(x,y)$, this 
implies the appearance of several components, corresponding to the various
positions of the time indices along the contour. For $G(x,y)$, the various contour components are written in a compact
manner by using the decomposition in terms of the correlators $G^>(x,y)\equiv \langle \varphi(x)\varphi(y)\rangle $ and
$G^<(x,y)\equiv \langle \varphi(y)\varphi(x)\rangle$, namely 
\begin{equation}
G(x,y)=\Theta_{\mathcal{C}}(x_0-y_0)G^>(x,y)+\Theta_{\mathcal{C}}(y_0-x_0)G^<(x,y).
\label{gdecomp}
\end{equation}
The $\Theta$-functions used here are defined along the contour
$\mathcal{C}$. A
similar decomposition can be written for the self-energy $\Sigma^{\rm nl}(x,y)$,
i.e.~
\begin{equation}
\Sigma^{\rm nl}(x,y)=\Theta_{\mathcal{C}}(x_0-y_0)\Sigma^>(x,y)+\Theta_{\mathcal{C}}(y_0-x_0)\Sigma^<(x,y).\label{sigmadecomp}
\end{equation}
 From \eref{gdecomp} we see that the
dynamics of the propagator is entirely described by the two complex functions $G^>$
and $G^<$. For the real scalar theory under consideration, these
functions satisfy
the property
$\left[G^>(x,y)\right]^{\star}=G^<(x,y)$, which leaves only one independent complex
function
describing the propagator dynamics. This can be
parametrized in terms of two real functions $F$ and $\rho$ according to
\begin{align}
G^>(x,y)&=F(x,y)-\frac{i}{2}\rho(x,y),\\
G^<(x,y)&=F(x,y)+\frac{i}{2}\rho(x,y).
\end{align}
The functions $F$ and $\rho$ correspond to the correlators
\begin{align}
F(x,y)&= \frac{1}{2}
\left[ G^>(x,y)+G^<(x,y)\right]=\half \left\langle \left\{\varphi(x),\varphi(y)\right\}
\right\rangle, \label{defF}\\
\rho(x,y)&=iG^>(x,y)-iG^<(x,y)=i\left\langle \left[ \varphi(x), \varphi(y)\right]
\right\rangle. \label{defrho}
\end{align}
The correlators $F(x,y)$ and $\rho(x,y)$ contain,
respectively, statistical
and spectral information about the system. 
They 
satisfy the symmetry properties
\begin{align}
F(x,y)&=F(y,x),\\
\rho(x,y)&=-\rho(y,x),
\end{align}
which make them very useful for numerical
implementation \cite{Aarts:2001qa}.
\par For the self-energy we introduce, in a similar fashion, the quantities 
\begin{align}
\Sigma^F(x,y)&=\frac{i}{2}
\left[ \Sigma^>(x,y)+\Sigma^<(x,y)\right], \\
\Sigma^{\rho}(x,y)&=\Sigma^<(x,y)-\Sigma^>(x,y),
\end{align}
which satisfy similar properties as their
propagator counterparts. For the case of the non-local part of the self-energy given by
\eref{sigmaNL}, these become
\begin{align}
\Sigma^F(x,y)&=\frac{\lambda^2}{2}\phi(x)\phi(y) \left[
F^2(x,y)-\frac{\rho^2(x,y)}{4}\right]+\frac{\lambda^2}{6}F(x,y) \left[
F^2(x,y)-\frac{3\rho^2(x,y)}{4}\right],\label{sigmaFth}\\
\Sigma^\rho(x,y)&=\lambda^2 \phi(x)\phi(y)\big[F(x,y)\rho(x,y)\big]
+\frac{\lambda^2}{6}\rho(x,y)\left[3F^2(x,y)-\frac{\rho^2(x,y)}{4}\right]\label{sigmarhoth}.    
\end{align}
\par In the study presented here, we shall focus on the dynamics of the statistical and spectral
correlators $F$ and $\rho$. Their equations of motion are determined from
\eref{eomG}
by using the decompositions \eref{gdecomp} and \eref{sigmadecomp}, as well as the definitions \eref{defF} and
\eref{defrho}. For the case $x_0>y_0$, one finds
\begin{align}
\left[ \partial_x^2+M^2(x)\right]F(x,y)&=\int_{0}^{x_0}dz_0 \int d^3
z\;\Sigma^{\rho}(x,z)F(z,y)-\int_{0}^{y_0}dz_0\int d^3z\;\Sigma^{F}(x,z)\rho(y,z),\label{eomFth}\\
\left[\partial_x^2+M^2(x)\right]\rho(x,y)&=\int_{y_0}^{x_0}dz_0\int
d^3z\;\Sigma^{\rho}(x,z)\rho(z,y),
\label{eomrhoth}
\end{align}
with
\be
M^2(x)=m^2+\frac{\lambda}{2}\phi(x)^2+\Sigma^{l}(x)=m^2+\frac{\lambda}{2}\phi(x)^2+\frac{\lambda}{2}F(x,x).
\label{M2}
\ee
With the same considerations as for the 2-point functions, the equation of
motion of the mean field $\phi(x)$ is found from \eref{eomphi} to be
\begin{equation}
\left[ \partial_x^2+M^2(x)-\frac{\lambda}{3}\phi(x)^2\right]\phi(x)=\int_0^{x_0}
dz_0\int d^3z\;
\widetilde{\Sigma}^\rho(x,z)\phi(z),
\label{eomphi2}
\end{equation}
where $\widetilde{\Sigma}^\rho(x,z)$ is the $\rho$-component of the ``sunset''
self-energy diagram, given by 
\be
\widetilde{\Sigma}^\rho(x,z)=-\frac{\lambda^2}{6}\rho(x,z)\left[3F(x,z)^2-\frac{\rho(x,z)^2}{4}\right].
\label{sigmamf}
\ee
This contribution derives from including the {\it second} of the 2PI diagrams in $\Phi$ (see
Eq.~\ref{3looptruncation}). Therefore it is present in both the two-loop 
and ``basketball'' approximations. The tilde
in $\tilde{\Sigma}$ is written to avoid any confusion with the self-energy $\Sigma$ entering in the equations of motion
for the propagator. In that case, the ``sunset'' diagram enters in the self-energy $\Sigma$ only in the ``basketball'' approximation. 
\par
With the self-energies $\Sigma^F$, $\Sigma^\rho$ and $\widetilde{\Sigma}^\rho$
given respectively by
(\ref{sigmaFth}),(\ref{sigmarhoth}) and \eref{sigmamf}, equations (\ref{eomFth}-\ref{eomphi2}) constitute a set of closed coupled evolution equations for the
correlators $F$ and $\rho$ and the mean field $\phi$. These equations are explicitly causal, i.e.~the
evolution of $F$, $\rho$ and $\phi$ is determined by the values of the
correlators and mean fields at previous times. The driving terms in the RHS of those
equations consist of nonlocal
``memory'' integrals that contain the information about the earlier stages of
the evolution. By specifying a complete set of initial conditions for $F$,
$\rho$ and $\phi$,
the equations of motion (\ref{eomFth}-\ref{eomphi2}) constitute an initial
value problem. We
perform a numerical analysis of the equations of motion in the next
section.
\par
We finish this section with the calculation of the energy density corresponding to the
truncations of the 2PI effective action. The energy density is
determined from the energy-momentum tensor component
$T^{00}$. It takes the form (see appendix
A)
\begin{align}
 T^{00}({\mathbf{x}},t) &=\half \Big[
 \partial_t\partial_{t^\prime}+\partial_{{\mathbf{x}}}\partial_{{\mathbf{x}}^\prime}+m^2\Big]\big(F({\mathbf{x}},t;{\mathbf{x}}^\prime,t^\prime)+\phi({\mathbf{x}},t)\phi({\mathbf{x}},t^\prime)\big)\bigg|_{\substack{
{\mathbf{x}}={\mathbf{x}}^\prime\\t=t^\prime
}}\nonumber \\
&\qquad +\frac{1}{4!}\lambda \phi(\mathbf{x},t)^4+\frac{1}{4}\lambda
\phi(\mathbf{x},t)^2F({\mathbf{x}},t;{\mathbf{x}},t)
-\frac{\delta \Phi}{\delta \zeta(x)}\Big|_{{\zeta = 1}}.
\label{energydensity}
\end{align}
Here $\zeta(x)$ is an auxiliary scale factor introduced in the coupling
constant as $\lambda \rightarrow \zeta(x)\lambda $.
For a given truncation, the energy density is obtained from \eref{energydensity} by the substitution $\Phi \rightarrow
\Phi_\ts{tr}$. In the ``basketball'' approximation, for instance, the energy density becomes
\begin{align}
T^{00} ({\mathbf{x}},t)&=\frac{1}{2}\left[\partial_t
\phi({\mathbf{x}},t) \right]^2+\frac{1}{2}\left[\partial_{{\mathbf{x}}}
\phi({\mathbf{x}},t)\cdot\partial_{{\mathbf{x}}}
\phi({\mathbf{x}},t) \right]+\half \partial_t \partial_{t^\prime}F({\mathbf{x}},t;{\mathbf{x}},t^\prime)\big|_{t=t^\prime}
+\half \partial_{{\mathbf{x}}}\cdot\partial_{\mathbf{y}}
F({\mathbf{x}},t;{\mathbf{x}},t)\big|_{{\mathbf{x}}=\mathbf{y}} \nonumber\\
&\ +\half m^2 \left[
\phi^2({\mathbf{x}},t)+F({\mathbf{x}},t;{\mathbf{x}},t)\right] 
+\frac{1}{4!}\lambda
\phi({\mathbf{x}},t)^4  +\frac{1}{4}\lambda F({\mathbf{x}},t;{\mathbf{x}},t)\phi({\mathbf{x}},t)^2
+\frac{1}{8}\lambda F({\mathbf{x}},t;{\mathbf{x}},t)^2\nonumber \\
&+\frac{\lambda^2}{6}\int_0^t dz_0\int
d^3z\;\phi({\mathbf{x}},t)\left[\frac{\rho({\mathbf{x}},t;{\mathbf{z}},z_0)^3}{4}-3\rho({\mathbf{x}},t;{\mathbf{z}},z_0)
F({\mathbf{x}},t;{\mathbf{z}},z_0)^2\right]\, \phi({\mathbf{z}},z_0)
\nonumber\\
&+\frac{\lambda^2}{6}\int_0^t dz_0\int
d^3z\;
\left[\frac{\rho({\mathbf{x}},t;{\mathbf{z}},z_0)^2}{4}-F({\mathbf{x}},t;{\mathbf{z}},z_0)^2\right]
F({\mathbf{x}},t;{\mathbf{z}},z_0)\rho({\mathbf{x}},t;{\mathbf{z}},z_0).
\label{energymomentumth}
\end{align}
It follows from translational invariance \cite{Baymconserve,Ivanov:1998nv}
(see also appendix A), that the energy density
\eref{energymomentumth} is exactly conserved in the evolution. 


\section{Numerical analysis and renormalization}
We study the non-equilibrium evolution of the correlators $F$ and $\rho$ and
the mean field $\phi$ by
solving numerically the equations of motion (\ref{eomFth}-\ref{eomphi2}), both in 
 the
symmetric and 
the
broken phase. 

\subsection{Numerical implementation}\label{numericalimple}
We shall consider the system to be discretized on a space-time
lattice with a finite spatial volume and spatially periodic boundary
conditions. The action of $\varphi^4$ theory on a space-time lattice is 
\be
S_\ts{lat}[\varphi]=a^3\,a_t\sum_{\mathbf{x},t} \left[ \half \left( \partial_t
\varphi(\mathbf{x},t)\right)^2-\half\sum_i \left(\partial_i
\varphi(\mathbf{x},t)\right)^2-\half
m_0^2\varphi(\mathbf{x},t)^2-\frac{1}{4!}\lambda_0
\varphi(\mathbf{x},t)^4 \right].
\ee
The lattice spacings $a$ and $a_t$ correspond to the spatial
and time directions, respectively. The derivatives stand for forward finite differences,
e.g. $\partial_t
\varphi(\mathbf{x},t)=(1/a_t)\left[\varphi(\mathbf{x},t+a_t)-\varphi(\mathbf{x},t)
\right]$. The spatial lattice volume is given in terms of the number of lattice sites
$N$ as $V=L^3=(Na)^3$. In the following we shall use lattice units ($a=1)$ and
write $dt=a_t/a$ for the dimensionless time-step. The mass $m_0$ and coupling
$\lambda_0$ are \emph{bare} parameters to be determined below. The lattice version of the
squared spatial momentum is given by
\be
k_\ts{lat}^2=\sum^3_{i=1}\left(2-2\cos \mathbf{k}_i\right),\quad
\mbox{with}\quad \mathbf{k}_i=\frac{2\pi n_i}{N},\quad
n_i=-\frac{N}{2}+1,\ldots,\frac{N}{2}\ (\mbox{for $N$ even}).
\ee
Plotting data as a function of $k_\ts{lat}^2$ corrects for a large part of the
lattice artifacts. 
\par The lattice provides a cutoff and regularizes the ultraviolet
divergent terms in the continuum limit, which 
 are 
to be dealt with by renormalization. The
continuum renormalization of $\Phi$-derivable approximations has been studied
in detail in \cite{vanHees:2001ik,VanHees:2001pf,Blaizot:2003an,Berges:2004hn,Cooper:2004rs}.
For our purpose it is enough to 
use 
an approximate renormalization
that ensures
that the relevant length scales in our simulations are larger than the
lattice spacing $a$. 
This is achieved by simply choosing 
the bare paremeters $m_0$ and $\lambda_0$
according to the one-loop formulas that relate 
them to the renormalized parameters.
The bare mass $m_0$ is given in terms
of the renormalized mass $m$ by (see also
\cite{Blaizot:2003an,Reinosa:2003qa,mythesis})
\be
m_0^2=m^2-\delta m^2,
\label{massren}
\ee
with the mass counterterm 
\be
\delta m^2=\frac{\lambda}{2a^2} I_1(am)-\frac{\lambda^2v^2}{2a^2}I_2(am).
\label{massren12}
\ee
The $I_1$ and $I_2$ in 
\eref{massren12} are dimensionless
integrals coming respectively from the one-loop ``leaf'' and ``eye''
diagrams in the self-energy $\Sigma$ at zero temperature. On the lattice, for continuous time and in
the infinite volume limit ($N\rightarrow \infty$), they are given by
\begin{align}
I_1(am)&=i \int_{-\pi/a}^{\pi/a}\frac{d^3k}{(2\pi)^3}\int
\frac{dk_0}{2\pi}
\frac{a^2}{k_0^2-a^{-2}k_\ts{lat}^2-m^2-i\epsilon}=\int_{-\pi}^{\pi}\frac{d^3k}{(2\pi)^3}\frac{1}{4\sqrt{\smash[b]{a^2m^2+k^2_{\ts{lat}}}}},\\
I_2(am)&=i \int_{-\pi/a}^{\pi/a}\frac{d^3k}{(2\pi)^3}\int
\frac{dk_0}{2\pi}\left(\frac{1}{k_0^2-a^{-2}k_\ts{lat}^2-m^2-i\epsilon}\right)^2=\int_{-\pi}^{\pi}\frac{d^3k}{(2\pi)^3}\frac{1}{8\sqrt{\smash[b]{\left(a^2m^2+k^2_{\ts{lat}}\right)^3}}}.
\end{align}
The bare coupling $\lambda_0$ can be determined from
the one-loop expression (see \cite{Blaizot:2003an,Reinosa:2003qa,mythesis})
\be
\frac{1}{\lambda_0}=\frac{1}{\lambda}-I_2(am).
\label{couplingren}
\ee
The renormalization conditions that define the renormalized mass and coupling via
Eqns.~\eref{massren} and \eref{couplingren} are such that they correspond to the values
of the two- and four-point vertex functions at vanishing external momenta.
For the values of the couplings ($\lambda =1,6$) and lattice
spacing ($0.5<am<1$) that we use in our simulations, the difference between $\lambda_0$ and $\lambda$ is less
than 10\%. In practice, we simply choose $\lambda_0$ as if it were the
renormalized coupling.
\par 
For the renormalization of the mass we use
\eref{massren}, which for the case of a spatial $N^3$ lattice is then given 
by
\be
m_0^2=\pm m^2-\frac{\lambda_0}{4a^2N^3}\sum_{\mathbf{k}}
\frac{1}{\sqrt{\smash[b]{a^2m^2+k^2_\ts{lat}}}}+\frac{\lambda_0^2v^2}{8a^2N^3}\sum_{\mathbf{k}}
\frac{1}{\sqrt{\smash[b]{\left(a^2m^2+k^2_\ts{lat}\right)^3}}}.
\label{massren2}
\ee
In practice, we conveniently choose a value for the renormalized mass (such
that $am<1$), 
which determines via
Eq. \eref{massren2} the bare mass that enters in the equations of motion for
the mean field and propagator. In our simulations we used $am=0.7$ and
$N=16$.
Given the input parameters $m_0$ and $\lambda_0$, the output physics is of
course not known precisely, it is determined by the $\Phi$-derived equations
of motion. 


\subsection{Initial conditions}

We 
specialize to a spatially homogeneous situation. In this case, 
$F(x,y)=F(t,t',{\bf x-y})$, $\rho(x,y)=\rho(t,t',{\bf x-y})$ 
so we can perform a Fourier transformation and study the propagator modes
$\rho_{\mathbf{k}}(t,t^\prime)$ and $F_{\mathbf{k}}(t,t^\prime)$. In addition, the
mean field depends only on time. To 
specify 
the time evolution, the equations of motion (\ref{eomFth}-\ref{eomphi2}) must be supplemented with initial conditions at $t=t^\prime=0$. 
These are given by the values and derivatives of $\rho$, $F$ and $\phi$ at initial time.
The initial conditions for $\rho$ follow from it being the expectation value of the commutator of two fields, which
implies
\be
\rho_{\mathbf{k}}(t,t)=0,\qquad \partial_t
\rho_{\mathbf{k}}(t,t^\prime)\big|_{t=t^\prime}=1, 
\label{rhoetr2}
\ee
Imposing the condition \eref{rhoetr2} at $t=t'=0$, it is preserved by the equations of motion.
\par
For the statistical correlator $F$, we choose initial conditions of the form
\begin{align}
\langle\{\varphi_{\bf k}(t),\varphi_{\bf-k}(t')\}\rangle|_{t=t'=0}=F_{\mathbf{k}}(t,t^\prime)\big|_{t=t^\prime=0}&=\frac{1}{\omega_{\mathbf{k}}}\left[n_{\mathbf{k}}+\half\right],\label{gaussianic1}\\
\langle\{\pi_{\bf k}(t),\varphi_{\bf -k}(t')\}\rangle|_{t=t'=0}=\partial_t F_{\mathbf{k}}(t,t^\prime)\big|_{t=t^\prime=0}&=0,\\
\langle\{\pi_{\bf k}(t)\pi_{\bf-k}(t')\}\rangle|_{t=t'=0}=\partial_t \partial_{t^\prime}
F_{\mathbf{k}}(t,t^\prime)\big|_{t=t^\prime=0}&=\omega_{\mathbf{k}}\left[
n_{\mathbf{k}}+\half \right] \label{gaussianic3}.
\end{align}
where $\pi_\mathbf{k}(t)=\partial_t \varphi_\mathbf{k}(t)$ are the 
conjugate 
field momenta, $n_{\mathbf{k}}$ is some distribution function and
$\omega_{\mathbf{k}}=\sqrt{\smash[b]{m_{\rm in}^2+\mathbf{k}^2}}$, 
with $m_{\rm in}$ to be specified shortly.
An initial condition of this form can be represented by a gaussian density 
matrix.
We will use the following cases for
the distribution function $n_{\mathbf{k}}$:
\begin{itemize}
\item[a)] \emph{Thermal}: The distribution function $n_{\mathbf{k}}$ corresponds to a
Bose-Einstein, at some initial temperature $T_\ts{in}$,
\begin{equation}
n_{\bf k}=\frac{1}{e^{(\omega_{\bf k}/T_{\rm in})}-1}.
\end{equation}
This also includes the ``vacuum'' initial condition of $T_\ts{in}=0$
(which is of course only an approximation to the vacuum state in the 
interacting theory). 
The input mass for all $T_\ts{in}$ is 
the renormalized 
$m_{\rm in} = m$ in the symmetric phase,
and $m_{\rm in}=\sqrt{2}m$ in the broken phase.
We expect these to be close to the zero-temperature particle masses, respectively in these phases.
\item[b)] \emph{Top-hat}: In this case, only modes with momenta within a range
${\mathbf{k}}_\ts{min}^2<{\mathbf{k}}^2<{\mathbf{k}}^2_\ts{max}$ are occupied.
The distribution function can be parametrized as
\be
n_{\mathbf{k}}=\eta\,\Theta({\mathbf{k}}^2_\ts{max}-{\mathbf{k}}^2)\Theta({\mathbf{k}}^2-{\mathbf{k}}^2_\ts{min}),
\label{top-hat}
\ee
where $\eta$ represents the occupancy of the excited modes. 
The input mass is again given by $m_{\rm in} = m$ 
(symmetric) and $m_{\rm in}=\sqrt{2}m$ (broken).
\end{itemize}
The mean field is initialized at $\phi=0$ (``symmetric phase'') and $\phi=v_\ts{tree}$, (the zero temperature ``broken phase''). Below, we will 
also
allow the mean field to be slightly displaced from these two, in order to study relaxation in a thermal background. 


\subsection{Observables}
As the system evolves in time, we expect the scattering processes to lead to
equilibration. The occupation numbers of the momentum modes are expected to
gradually approach a Bose-Einstein distribution,
provided the coupling is not too strong.
The statistical information about the evolving system can be extracted from
the equal-time correlation function $F_{\mathbf{k}}(t,t)$. We can use
$F_{\mathbf{k}}$ to define a quasiparticle distribution function and frequencies as
\cite{Aarts:1999zn,Salle:2000hd,Aarts:2001qa,Salle:2002fu,Skullerud:2003ki}
\begin{align}
n_{\mathbf{k}}(t)+\half&=c_\mathbf{k}\sqrt{\partial_t \partial_{t^\prime}
F_{\mathbf{k}}(t,t^\prime)\big|_{t=t^\prime}\,F_{\mathbf{k}}(t,t)},\label{particlenumber}\\
\omega_{\mathbf{p}}(t)&=\sqrt{\frac{\partial_t \partial_{t^\prime}
F_{\mathbf{k}}(t,t^\prime)\big|_{t=t^\prime}}{F_{\mathbf{k}}(t,t)}}.
\label{particleenergy}
\end{align}
The correction $c_\mathbf{k}$ diminishes errors associated with the time
discretization on the lattice. It is given by 
\cite{Salle:2002fu}
\be
c_\mathbf{k}=\sqrt{1-\frac{1}{4}dt^2\omega_\mathbf{k}^2}.
\label{ck}
\ee
Both definitions
(\ref{particlenumber}) and \eref{particleenergy} are valid for a free field
system in equilibrium, and have proven to be very useful in interacting
theories out of equilibrium as well
\cite{Salle:2000hd,Salle:2000jb,Skullerud:2003ki,Salle:2003ju}. From the studies in
1+1 and 2+1 dimensions \cite{Aarts:2001qa,Juchem:2003bi}, we expect the
system to exhibit a quasiparticle structure before reaching thermal
equilibrium. The definitions
\eref{particlenumber} and \eref{particleenergy} can be used to monitor the
evolution of the system towards such a quasiparticle-like state, and eventually to equilibrium.
\par
Once the system is close to equilibrium, we can read from \eref{particleenergy} the effective
quasiparticle mass $m_\ts{eff}(t)$ by comparing it to the dispersion relation 
\be
\omega^2_{\mathbf{k}}(t)=c^2(t)\left(m_\ts{eff}^2(t)+{\mathbf{k}}^2\right),
\label{disprelfit}
\ee
where the factor $c(t)$ is a measure of an effective 
speed of light or an inverse refractive index.
A temperature $T_\ts{eff}(t)$ and chemical
potential $\mu_\ts{eff}(t)$ can be determined by fitting the occupation number \eref{particlenumber} to a Bose-Einstein distribution
\be
n_{\mathbf{p}}(t)=\frac{1}{e^{\left[\omega_{\mathbf{p}}(t)-\mu_\ts{eff}(t)\right]/T_\ts{eff}(t)}-1}.
\ee
using
\be
\ln
\left(1+\frac{1}{n_{\mathbf{p}}}\right)=\frac{1}{T_\ts{eff}}\omega_{\mathbf{p}}-\frac{\mu_\ts{eff}}{T_\ts{eff}}.
\label{Bosefit}
\ee

\par
We also keep track of the `memory kernels' in the equations of motion
(\ref{eomFth}-\ref{eomphi2}), i.e.~the self-energies $\Sigma^F(t,t^\prime)$, $\Sigma^\rho(t,t^\prime)$ and $\tilde{\Sigma}^\rho(t,t^\prime)$, which can be compared with perturbative
estimates. 
Limits on computer resources (memory and CPU time) requires us to cut the memory kernels and thus keep only some
finite range backwards in time (i.e.~$\Sigma(t,t^\prime)\rightarrow 0\
\mbox{for}\ |t-t^\prime|>t_\ts{cut}$). The size of the self-energies helps
us determine whether the cut was late enough for the discarded memory
integrals to be negligible. A way to judge whether the discarded memory was
indeed unimportant for the dynamics is to verify that the total energy density
\eref{energymomentumth} is conserved. Monitoring the evolution of the energy
density \eref{energymomentumth}, one finds that at very late times, the
effect of the memory cut shows up as a very slow drift in the energy. In all
the runs presented here, the energy is conserved to within 2\%. For smaller
lattices we checked that later memory cuts make the drift smaller. We found no such drift if the whole kernel was kept. 


\subsection{Symmetric phase: equilibration}

We first consider the evolution of the system in the symmetric phase.
The simulations are performed on a lattice with $N^3=16^3$ sites, lattice
spacing $am=0.7$, time-step
$dt=0.1$ and coupling\footnote{Recall that we neglect the difference between $\lambda$ and $\lambda_0$.} 
$\lambda=6$. 
The memory kernel is cut off at
$mt_\ts{cut}=28$ unless otherwise specified. In the
following, all
quantities shall be expressed in renormalized mass units, i.e.~units of $m$. 
\par
For the initial conditions of the propagators we shall take: 
\begin{itemize}
\item Thermal, with $T_\ts{in}/m=1.36,1.43,1.93,2.86$.
\item Top-hat 1 (T1),  with ${\mathbf{k}}^2_\ts{min}/m^2=2.04$,
 ${\mathbf{k}}^2_\ts{max}/m^2=6.12$ and $\eta=2$.
 \item Top-hat 2 (T2), with ${\mathbf{k}}^2_\ts{min}/m^2=0$,
${\mathbf{k}}^2_\ts{max}/m^2=5.71$ and $\eta=1.85$.
\item Top-hat 3 (T3), with ${\mathbf{k}}^2_\ts{min}/m^2=6.12$,
 ${\mathbf{k}}^2_\ts{max}/m^2=8.16$ and $\eta=1.6$.
 \end{itemize}
The three top-hat initial conditions have the same initial energy. 

In a quasi-particle picture, we can introduce the total particle number 
density
\begin{equation}
	n_{\rm tot}=\frac{N_\ts{tot}}{V}=
\begin{cases}
\int \frac{d^{3}k}{(2\pi)^{3}}n_{\bf k},& \text{ 
for infinite volume in the continuum,} \\
\frac{1}{N^3 a^3}\sum_{\bf k}n_{\bf k},& \text{on the lattice.}
\end{cases}
\end{equation} 
The three top-hat initial conditions T1-T3 do not have the same total number of particles, although T1 and T2 are fairly close to each other. More about this below.
\par
For the initial mean field we take $\phi(0)=0$. In this case, the Hartree and the two-loop approximations are
identical. In the following we consider the evolution in both the Hartree and the ``basketball'' approximations.
\par 
\begin{figure}[h]
\centerline{
\epsfig{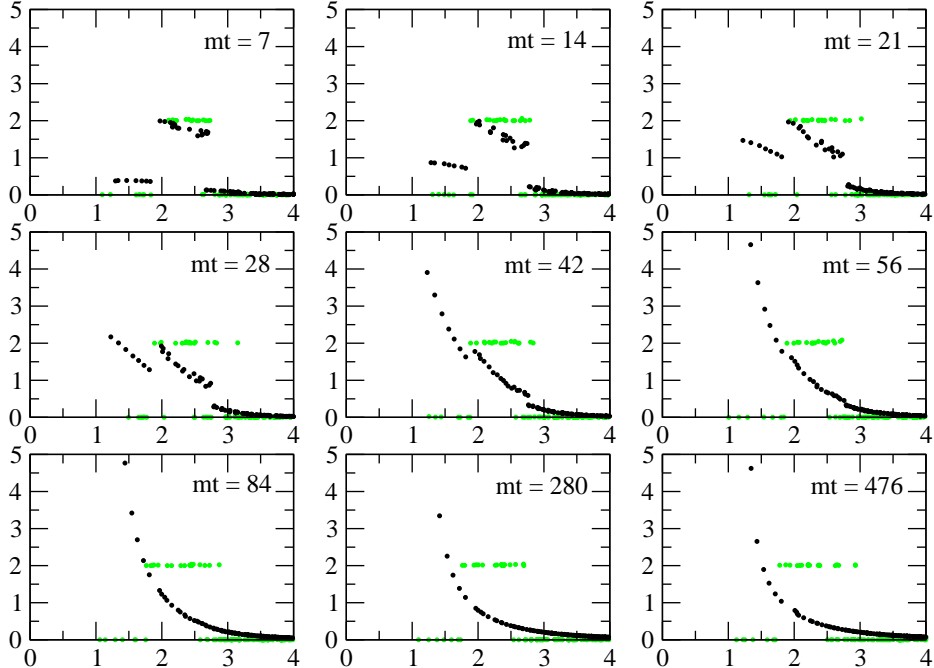}
}
\caption{Evolution in time of the occupation numbers $n_{\mathbf{k}}$
\textit{vs.~}$\omega_{\mathbf{k}}$, for the T1 initial condition. We display the results of the ``basketball''
(black 
dots) and the Hartree approximation
(green/grey 
dots).}
\label{fig:Waterloosunset}
\end{figure}
In Fig.~\ref{fig:Waterloosunset}, we show the evolution of $n_{\mathbf{k}}$ versus $\omega_{\mathbf{k}}$, starting from the T1 initial condition. The Hartree approximation (black) is compared with the ``basketball'' approximation (green/grey). In the former case, there is no equilibration. For the ``basketball'' case, we observe that the energy in the excited modes is distributed via scattering.
As we shall see this leads eventually to a thermal distribution.
\par
\begin{figure}
\centerline{
\epsfig{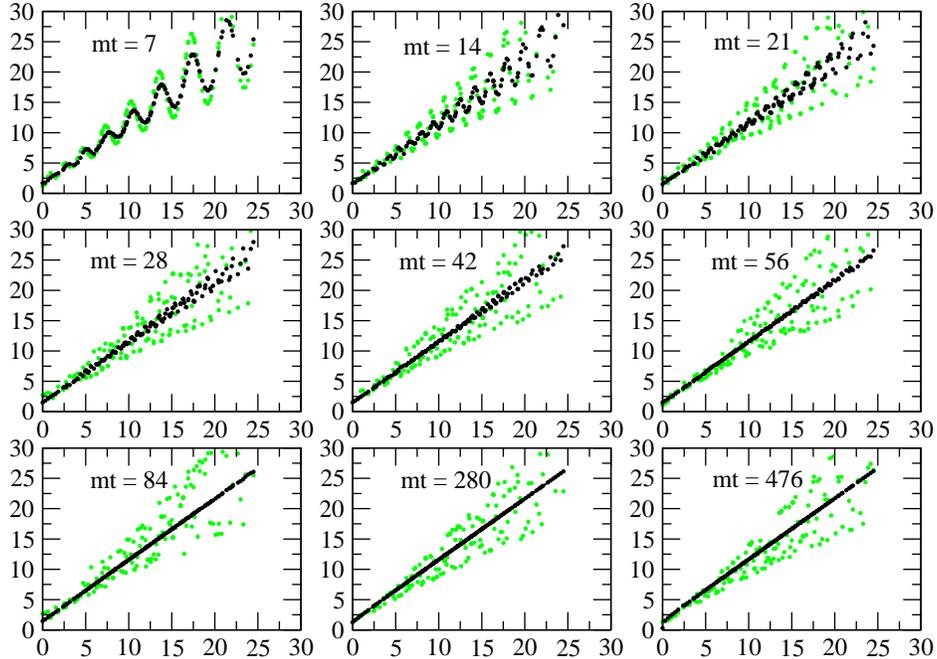}
}
\caption{Time evolution of the dispersion relation ($\omega_{\mathbf{k}}^2$
vs.~${\mathbf{k}}^2$) in the symmetric phase, starting from the T1 initial conditions, in the ``basketball'' (
 black
dots) and Hartree approximations (green/grey
dots).}
\label{fig:omkSunS}
\end{figure}

In Fig.~\ref{fig:omkSunS} we follow the evolution of the dispersion relation with the same initial condition, again comparing Hartree to ``basketball''. Notice the 
oscillating pattern early on 
in
both cases. 
In the ``basketball'' approximation the modes eventually relax to a perfect straight line. It turns out that at the couplings and energies used here, the coefficient $c^{2}(t)$ is equal to 1 up to well within one percent. We will therefore assume it to be 1 in the following. Although we do not show it here, we found that larger coupling and large energy density results in a faster evolution towards this quasi-particle state. 
\par
Judging by eye, Figs.~\ref{fig:Waterloosunset} and \ref{fig:omkSunS} suggest that already at times $mt=56$ to $84$,
the system behaves as aproximately thermal, for T1 initial conditions.
Still, this is presumably much later than a pre-thermalization time
based on the the equation of state, as studied in \cite{Berges:2004ce}.
However, it does not mean that the memory of the initial conditions 
in the particle distribution is already lost by times $mt \gtrsim 84$. 
\par
\begin{figure}[h]
\centerline{
\epsfig{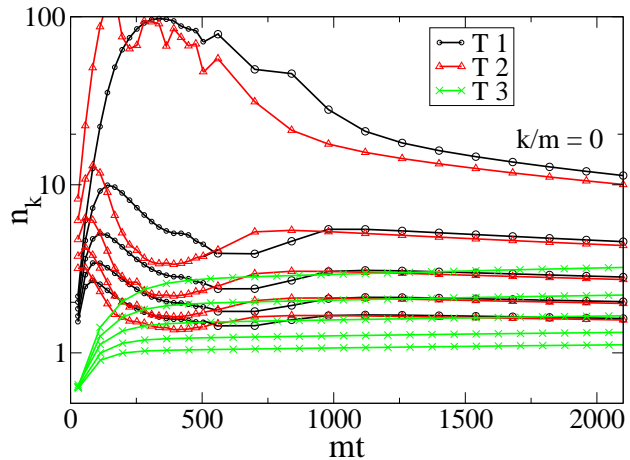}
}
\caption{Evolution of individual modes for a T1 
, T2 
 and T3 
 initial condition with same energy density.}
\label{fig:modeasymptote}
\end{figure}

It was remarked already in 
the studies in 1+1 \cite{Berges:2000ur} and 2+1
dimensions \cite{Juchem:2003bi} that the final state depends only on the
energy density (at a given coupling). Indeed, it was found that the limit
distribution function $n_\mathbf{k}$ corresponds to a Bose-Einstein,
characterized by just one parameter, the temperature. Fig.~\ref{fig:modeasymptote} shows the evolution of
individual modes when starting from the T1, T2 and T3 initial conditions, which have the same energy density. For T1 and
T2 we see that the modes 
approach a common final value. It seems reasonable to call this stage
\emph{kinetic equilibration}, as the kinetic energy is redistributed over the modes to reach a Bose-Einstein
distribution. However, as we will see below, the total number of particles 
is not adjusting as fast and it still
remembers the initial state by the time kinetic equilibration is completed.

As mentioned before, the initial condition T3 has not only a different initial spectrum, but also a different total
number of particles. It also reaches kinetic equilibration, but with a different kinetically equilibrated state.
\par
\begin{figure}[h]
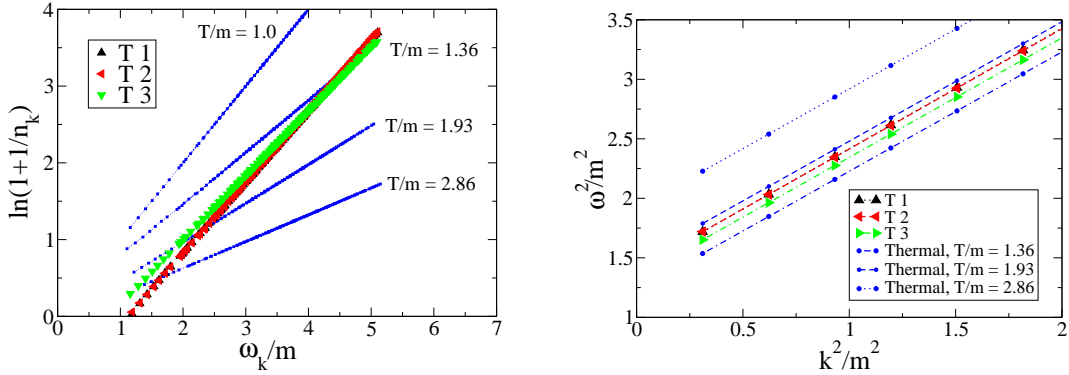

\centerline{
\epsfig{file=pictures/P4_nk_sym_early_ooe.eps,width=.4\textwidth,clip}
$\qquad$
\epsfig{file=pictures/P4_disp_sym_early_ooe.eps,width=.4\textwidth,clip}
}
\caption{Left: occupation numbers $\ln (1+1/n_\mathbf{k})$ at intermediate
times $mt=1000$ for initial T1, T2 and T3 and a set of thermal initial states. Right: The dispersion
relation for the same cases. The intercepts give the effective masses squared.}
\label{fig:intermediatecomparison}
\end{figure}

At intermediate times ($mt\approx 1000$) kinetic equilibration has taken place, and we can compare the distribution
functions and dispersion relations, Fig.~\ref{fig:intermediatecomparison}. T1 and T2 have equilibrated to almost
identical Bose-Einstein distribution functions, parametrized by an effective mass, an effective temperature and an
effective chemical potential. T3 has reached a different Bose-Einstein with a different temperature and chemical
potential, and a slightly different effective mass. We have included a number of thermal initial conditions for
comparison. By construction, these have no initial chemical potential and remain so to a very good approximation.

Whereas kinetic equilibration can be the result of simple $2 \leftrightarrow 2$ scattering, chemical equilibration, which changes the total particle number, happens through $1\leftrightarrow 3$, $2\leftrightarrow 4$ and higher order processes. These are included due to the resummations performed by the $\Phi$-derivable approximation into the ``sunset'' self-energy diagram. Approaches that only take into account on-shell scattering, such as the Boltzmann equation with only binary collisions $2 \leftrightarrow 2$, cannot account
for chemical equilibration.
What we see is that kinetic equilibration {\em including memory loss} happens on a time scale of about $500-1000/m$, whereas chemical equilibration is a much slower process. Effectively, there is a chemical potential in the initial stages, 
causing
initial conditions with different $N_{\rm tot}$ to relax to different intermediate kinetically equilibrated states. 

\begin{figure}
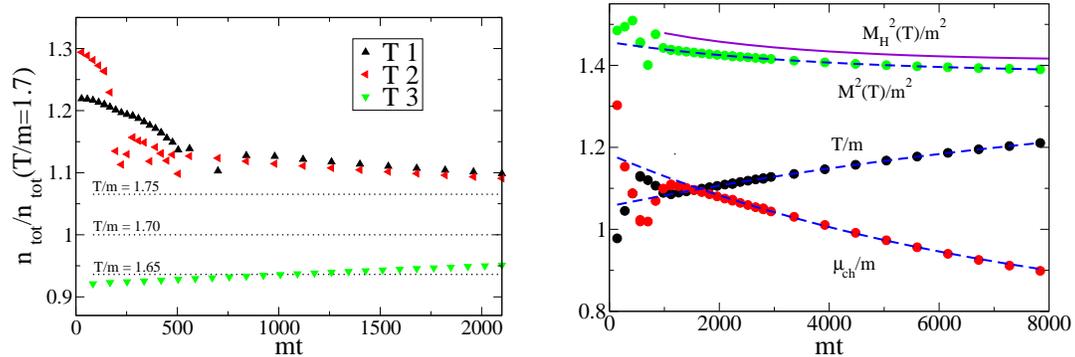

\centerline{
\epsfig{file=pictures/P5_Ntot_evol_ooe.eps,width=.4\textwidth,clip}
$\qquad$
\epsfig{file=pictures/P5_muT_evol_sym.eps,width=.4\textwidth,clip}}
\caption{Left: Total particle number density 
$n_{\rm tot}$ 
versus
time for the initial conditions T1, T2 and T3. 
The dotted lines are $n_{\rm tot}$ for Bose-Einstein distributions at different temperatures. Right: Evolution of the effective mass (squared),
temperature and chemical potential for late times starting from the T1 initial condition, with exponential fits
(dashed
lines). The Hartree estimate for the mass, 
(\ref{M2}), and the
solution of the gap equation \eref{latticegapeq}, 
are also 
shown.}
\label{fig:latest}
\end{figure}
We illustrate this point in Fig.~\ref{fig:latest}, left-hand plot, where we show the evolution $N_{\rm tot}$ for
the initial conditions T1-T3. For comparison, we also include the $N_{\rm tot}$ of Bose-Einstein distributions at
various temperatures. In the right-hand plot we follow the evolution of the effective mass, temperature and chemical
potential for the T1 case. The time evolution can be well reproduced by exponential fits of the form $a_{i}+b_{i}\exp{(-\gamma_{i}t)}$ (the dashed lines in the plot), suggesting an asymptotic temperature of around $T/m=1.36$. Also, within a factor of two, fits to the three quantities all suggest an equilibration time of around $\gamma_{i}^{-1}\simeq 10^{4}/m$. Chemical equilibration is a full order of magnitude slower than kinetic equilibration in this system.
Comparing with the study in \cite{Juchem:2003bi}, it appears that chemical equilibration is much slower in 3+1 than in 2+1 dimensions.
The fit to the chemical potential is not as good as 
to the effective temperature or mass,
and it also predicts a non-zero asymptotic chemical potential ($\mu/m=0.7$). This is consistent with our above-mentioned interpretation that the system is in a prethermalized stage \cite{Berges:2004ce} for which an exponential extrapolation of the evolution of $T$ and $\mu$ does
not necessarily yield the actual asymptotic values. 
\par
As an aside we compare the observed mass with an 
estimate that results from the Hartree approximation.
At a given time $t$, the finite gap equation for the Hartree effective mass
reads
\be
M_\ts{H}^2(t)=m^2+\Sigma^{\ts{l}}(t)-\delta m^2=m^2+\frac{\lambda}{2}\int
\frac{d^3k}{(2\pi)^3}F_\mathbf{k}(t,t)-\delta m^2.
\label{2loopren}
\ee
The mass counterterm $\delta m^2$ is given by the vacuum part of the ``leaf''
self-energy diagram, as described previously.
For the correlator $F(t,t)$ in Eq.~\eref{2loopren} we take the same form as a free quasiparticle gas in equilibrium, i.e.
\be
F_\mathbf{k}(t,t)=\frac{1}{\omega_\mathbf{k}(t)}\left[n_\mathbf{k}(t)+\half \right].
\ee
Here $n_\mathbf{k}(t)$ is a Bose-Einstein distribution function with the
temperature and chemical potential 
obtained from the
simulations at time $t$, and $\omega_\mathbf{k}(t)$ is 
here defined
in terms of the effective mass as 
$\omega_\mathbf{k}(t)=\sqrt{\smash[b]{\mathbf{k}^2+M_\ts{H}(t)^2}}$.
The result for the Hartree effective mass $M_\ts{H}$ is then determined by the self-consistent gap equation
\be
M_\ts{H}^2(t;T,\mu)=m^2+\frac{\lambda}{2}\int
\frac{d^3k}{(2\pi)^3}\frac{n_{\mathbf{k}}(t;T,\mu)}{\omega_{\mathbf{k}}(t)}=m^2+\frac{\lambda
T^2}{4\pi^2}\int_{M_\tinys{H}/T}^{\infty}dx
\frac{\sqrt{x^2-\frac{M_\tinys{H}^2}{T^2}}}{e^{\left(x-\frac{\mu}{T}\right)}-1}.
\label{gapeq}
\ee
To compare with the numerical result we use the lattice analog of the gap equation \eref{gapeq}, i.e.
\be
M_\ts{H}^2(t;T,\mu)=m^2+\frac{\lambda}{2}\frac{1}{(Na)^3}\sum_{\mathbf{k}}
\frac{n_{\mathbf{k}}(t;T,\mu)}{\omega_{\mathbf{k}}(t)}.
\label{latticegapeq}
\ee
For the case of the evolution starting from the T1 initial
condition, the lattice Hartree mass is shown in Fig.~\ref{fig:latest}.
We see that it is slightly higher than the effective mass obtained in the simulation with the ``basketball''
approximation. At least in this case, the contribution from the ``sunset'' diagram to the mass appears to be small relative to the Hartree case.


\subsection{Symmetric phase: Damping and the spectral function}\label{sec:symmetricphase}

\subsubsection{Mean field damping}
We now consider a situation already in (or close to) thermal equilibrium, where the mean field is
slightly displaced from its equilibrium value $\phi=0$. This allows us to
study the response of the system to small perturbations. In this case, the mean field
evolution can be studied by linearizing the equation of motion \eref{eomphi2}
around the equilibrium value. For homogeneous fields this leads to
\be
\overset{..}{\phi}(t)+M^2(T,t)\phi(t)+\int_0^t dt^\prime\,
\tilde{\Sigma}^\rho_\mathbf{0}(t,t^\prime)\,\phi(t^\prime)=0.
\label{linearizedeomphi}
\ee
with $\tilde{\Sigma}^\rho_\mathbf{0}(t,t^\prime)$ the zero momentum
mode of the ``sunset'' self-energy and $M(T,t)$ given by \eref{M2}. 
Close enough to equilibrium we may assume time translation invariance, such that
$\tilde{\Sigma}^\rho_\mathbf{0}(t,t^\prime)$ depends only on $t-t'$ and $M(T,t)$
is constant.
The equation \eref{linearizedeomphi} can 
then 
be solved by a
Laplace transform in the time coordinate \cite{Boyanovsky:1996xx}. Taking as
initial conditions for the mean field $\phi(0)=\phi_i$ and
$\overset{.}{\phi}(0)=0$, the solution to the linearized equation of motion
\eref{linearizedeomphi} can be written as \cite{Boyanovsky:1995me,Boyanovsky:1996xx}
\be
\phi(t)=\frac{2\phi_i}{\pi}\int_0^\infty d\omega \frac{\omega\,
\mbox{Im}\tilde{\Sigma}^R_\mathbf{0}(\omega)\,\cos (\omega t)}{\left[\omega^2-M^2-\mbox{Re}\tilde{\Sigma}^R_\mathbf{0}(\omega)\right]^2+\mbox{Im}\tilde{\Sigma}^R_\mathbf{0}(\omega)^2},
\label{evphit}
\ee
where $\mbox{Re}\tilde{\Sigma}^R$ and $\mbox{Im}\tilde{\Sigma}^R$ correspond
respectively to the real and imaginary part of the retarded self-energy, given by 
$\tilde{\Sigma}^R(x,y)=\Theta(x_0-y_0)\tilde{\Sigma}^\rho(x,y)$.
For weak coupling there is a narrow resonance at
$\omega=M_\ts{eff}$, with 
$M_\ts{eff}^2\equiv
M^2+\mbox{Re}\tilde{\Sigma}^R_\mathbf{0}(\omega)$. To a good approximation, one finds that for
short times the evolution is given by \cite{Boyanovsky:1995em}
\be
\phi(t)\approx \phi_i Z\, e^{-\gamma t}\cos\left( M_\ts{eff}
t-\alpha\right),
\label{evolmf}
\ee
with
\begin{align}
Z&=\bigg[1-\frac{\partial \mbox{Re}\tilde{\Sigma}^R_\mathbf{0}(M_\ts{eff})}{\partial
M^2_\ts{eff}} \bigg]^{-1},\\
\alpha&= \frac{\partial \mbox{Im}\tilde{\Sigma}^R_\mathbf{0}(M_\ts{eff})}{\partial
M^2_\ts{eff}} , \\
\gamma&= Z \frac{\mbox{Im}\tilde{\Sigma}^R_\mathbf{0}(M_\ts{eff})}{M_\ts{eff}} \label{osdrformal}.
\end{align}
The parameter $\gamma$ corresponds to the on-shell damping rate. For weak
enough couplings one can approximate $Z\approx 1$ and $M_\ts{eff}=M$ for the
calculation of $\gamma$. From \eref{osdrformal} we see that the damping rate is determined by the imaginary part of
$\tilde{\Sigma}^R$, which corresponds to the ``sunset'' self-energy diagram. In the context of perturbation theory,
$\mbox{Im}\,\tilde{\Sigma}^R$ can be calculated analytically and the damping
rate is found to be \cite{Wang:1996qf}
\be
\gamma(M)=\frac{\lambda^2 T^2}{128 \pi^3 M}\,\mbox{Li}_2(e^{-M/T})
\label{dampingrate}
\ee
where $\mbox{Li}_2(z)$ is the second polylogarithmic function, defined by Spence's integral
\be
\mbox{Li}_2(z)=-\int_0^{z} dw
\frac{\ln\left({1-w}\right)}{w}.
\ee
For temperatures $T \gg m$, the damping rate follows from the expression of
the high-temperature screening mass \cite{Arnold:1992rz}
\be
M^2=m^2+\frac{\lambda
T^2}{24}-\frac{\lambda}{8\pi}M\,T+\mathcal{O}\left(\lambda
M\,\mbox{ln}\frac{M^2}{T^2}\right).
\label{thermalmass}
\ee
In the limit of very
weak coupling and high temperature one obtains for the damping rate the compact result\footnote{This approximation for
the damping rate is often used in the literature. For the values of the coupling $\lambda=1,6$ used in the numerical
analysis presented in this paper, however, the approximation
\eref{highTdr} to \eref{dampingrate} is not valid, even for
high temperatures.}
  \cite{Jeon:1992kk,Wang:1996qf}
 \be
\gamma\Big|_{T \gg m,\,\lambda \ll 1}=\frac{\lambda^2 T^2}{768 \pi M}.
\label{highTdr}
\ee
\par
\begin{figure}
\centerline{
\epsfig{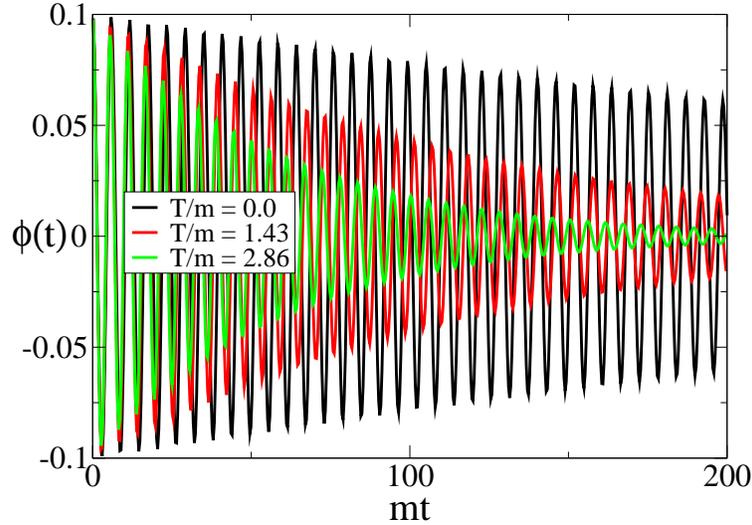}}
\caption{Time evolution $\phi(t)$ of the initially displaced mean field
$\phi_{i}=0.142$ for the cases $T_\ts{in}/m=0,1.43,2.86$, in the ``basketball'' approximation.}
\label{fig:evolmf}
\end{figure}
\begin{figure}
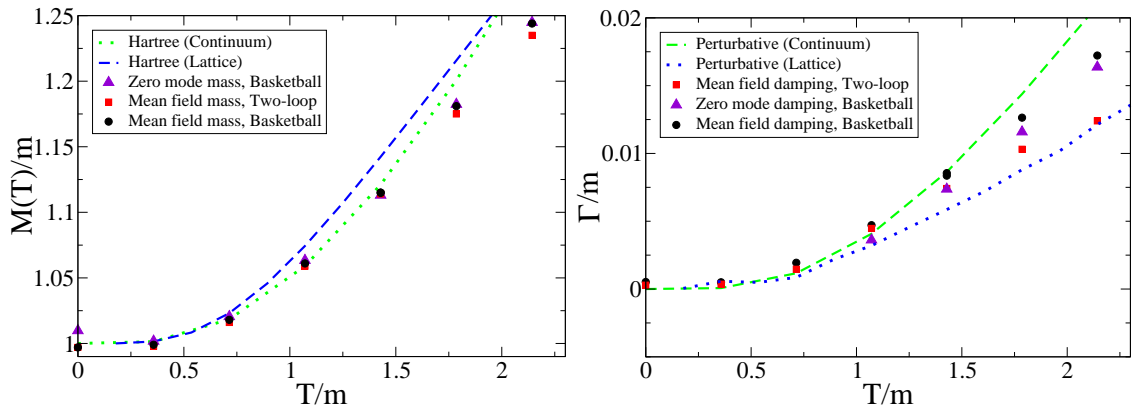

\begin{center}
\epsfig{file=pictures/P8_damp_mass_all.eps,width=0.45\textwidth,clip}
\epsfig{file=pictures/P8_damp_gamma_all.eps,width=0.45\textwidth,clip}
\end{center}
\caption{Left: Effective masses from the analysis of the evolution of the mean field (in the two-loop and ``basketball'' aproximations) and spectral function 
zero-mode (``basketball'' case). Hartree estimates,
both in the continuum (dotted line) and on the lattice (dashed line), are 
also shown.
Right: Damping rates. 
}
\label{fig:plotdampall}
\end{figure}

In the numerical simulations the mean field is initially displaced to the value $\phi_i/m=0.142$. For such a small perturbation, the
mean field is expected to perform a damped oscillation of the type
\eref{evolmf}. We fit the evolution of the mean field with Eq.~\eref{evolmf}, which allows us to extract the effective
mass $M_\ts{eff}$ and the damping rate $\gamma$. These will depend on the strength of the coupling and the temperature.
The behavior of the mean field is studied in a thermal bath at temperatures $T_\ts{in}/m=0,1.43,2.86$ for both the
two-loop and ``basketball'' approximations. For the ``basketball'' case, the mean field evolution is shown in Fig.~\ref{fig:evolmf}. 
As mentioned earlier, the damping of the mean field is present in both the two-loop and ``basketball''
approximations. The difference between the two 
cases 
lies 
in the fact 
that the correlators $F$ and $\rho$ evolve quite
differently. For the two-loop case, the equations of motion for $F$ and $\rho$ contain almost no damping, since the only potential
contribution to damping is in the ``eye'' diagram, which is proportional to $\phi^2$, and thus tiny for
$\phi\approx 0$. For the ``basketball'' case, however, the equations of motion for $F$ and $\rho$ contain damping
through the ``sunset'' diagram. 
The differences in the evolution 
of the correlators for the two-loop and the ``basketball''
approximations enters as a higher-order effect in the evolution of the mean field. 
In particular, 
this
may lead 
to different
effective masses and mean field damping rates. We 
show 
these differences for various temperatures in
Fig.~\ref{fig:plotdampall}, where the results for the two-loop (squares) and ``basketball'' (large dots) are plotted. 
\par
For comparison we 
evaluated 
the perturbative result (\ref{dampingrate}), 
using the Hartree mass \eref{gapeq} 
for $M$.
To see the finite-volume and discretization effects we also did the analytical
computation 
on a spatial lattice. The Hartree mass is in this case given by \eref{latticegapeq}. 
In finite volume, the discreteness of the momenta leads to complications 
in the calculation of the damping rate,
which we dealt with along the lines presented in 
\cite{Salle:2000jf}\footnote{For example,
equations like (\ref{imeyeintegral}) in the appendix do not make sense anymore.
We evaluated the 
frequency integral in the solution for the linearized equation of motion 
\eref{evphit} for $\phi(t)$, using a finite ``$i\epsilon$'' in the retarded 
sunset self-energy on the lattice.}.  
The perturbative results for the mass and damping rate are also presented in Fig.~\ref{fig:plotdampall}.
As we can see from the mass plot (left), the correction to the mass coming from the ``basketball'' approximation is small
relative to the Hartree case. In the damping plot (right) we observe that the damping rate obtained from the numerical
analysis of both approximations is substantially larger (about 20-40\%) than the perturbative result (on the lattice). The continuum and
the lattice perturbative results begin to differ 
around $T/m \gtrsim 1$ due to cut-off effects. 
\par
%

\subsubsection{Propagator damping and spectral function}
\par
Damping in the propagator 
can be elegantly 
phrased in terms of the spectral function $\rho_{\bf k}(t,t')$. 
In a situation close to thermal equilibrium 
we expect it to be time translation invariant and 
in a narrow-width approximation be
given by
\begin{align}
\rho_{\bf k}(t,t')=\frac{1}{\omega_\mathbf{k}} e^{-\gamma_{\bf k} |t-t'|}\sin\left[\omega_{\bf k}(t-t')\right],
\label{spectrform}
\end{align}
To study the approach to equilibrium of the spectral funcion, it is 
useful to perform a Wigner transformation in
terms of the 
mean time $\mathcal{T}=(t+t^\prime)/2$ and relative time $\tau=t-t^\prime$. This can be written as
\begin{align}
\rho_{\bf k}(\omega,\mathcal{T})=2i\int_{0}^{2\mathcal{T}}d\tau \sin(\omega \tau)\rho_{\bf
k}(\mathcal{T}+\tau/2,\mathcal{T}-\tau/2).
\label{wigner}
\end{align}
Since we are solving the equations of motion in a finite time and keep information only as far back as the memory kernel, we 
have a cut-off 
in the integral of \eref{wigner} as
\be
\rho_{\bf k}(\omega,\mathcal{T})\approx \rho_{\bf k}(\omega,\mathcal{T};t_\ts{cut})=2i\int_{0}^{t_\ts{cut}}d\tau \sin(\omega \tau)\rho_{\bf k}(\mathcal{T}+\tau/2,\mathcal{T}-\tau/2).
\label{modifiedwigner}
\ee
If the system is sufficiently close to equilibrium, time translation invariance should be a good approximation, 
$\rho_{\bf k}(t,t')\approx \rho_{\bf k}(t-t')$, and
\begin{align}
	\rho_{\bf k}(\omega,\mathcal{T};t_\ts{cut})\approx 2i\int_{0}^{t_\ts{cut}}d\tau\,\sin{(\omega\tau)}\,\rho_{\bf
	k}(\mathcal{T},\mathcal{T}-\tau).
\label{wignertrans}
\end{align}
With this approximation the 
integrand
in \eref{wignertrans} runs from $\rho_\mathbf{k}(\mathcal{T},\mathcal{T})$ to 
$\rho_\mathbf{k}(\mathcal{T},\mathcal{T}-t_\ts{cut})$, 
which 
is
convenient for numerical purposes. In the following we shall make use of this approximate Wigner transform.  
\par
\begin{figure}
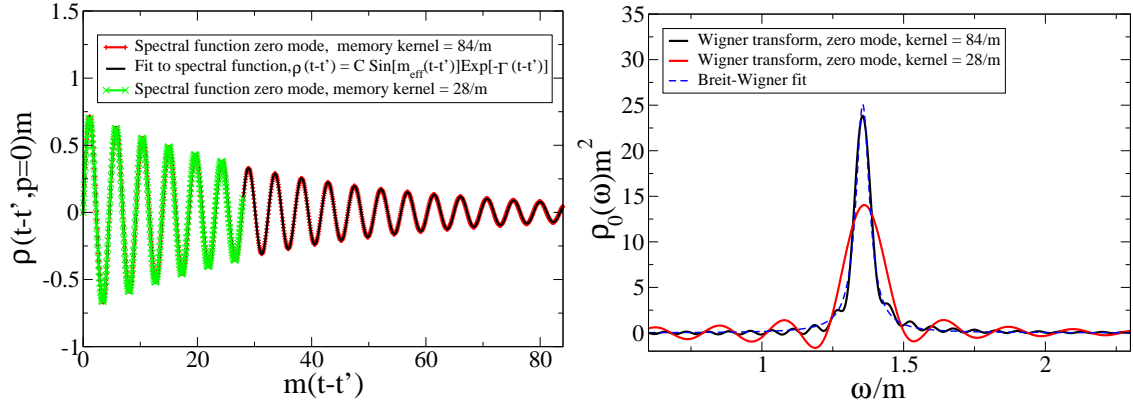

\begin{center}
\epsfig{file=pictures/P8_prop_damp_ex_sym.eps,width=0.45\textwidth,clip}
\epsfig{file=pictures/P8_prop_wign_ex_sym.eps,width=0.45\textwidth,clip}
\end{center}
\caption{Left: The evolution of the spectral function $\rho_\mathbf{0}(t,t^\prime)$ starting from a thermal initial
condition with $T/m=2.86$ and with coupling $\lambda=6$. The result with a short and a long kernel is plotted.
Overlaid a fit to the form (\ref{spectrform}). Right: The approximate Wigner transform
$\rho_\mathbf{0}(\omega)$ at time $m\mathcal{T}=200$. A Breit-Wigner fit to the Wigner transform with kernel $84/m$ is
included (dashed line).}
\label{fig:plotdampprop}
\end{figure}

The approximation \eref{wignertrans} is valid provided $\mathcal{T}$ is large enough so that the system is close to
thermal equilibrium. In that case $\rho_{\bf
k}(\omega_{\bf k},\mathcal{T})$ should be well approximated by a Breit-Wigner form
\begin{align}
\rho_{\bf k}(\omega,\mathcal{T})\approx \frac{4\omega(\gamma_{\bf k}/2)}{(\omega^{2}-\omega_{\bf
k}^{2})^{2}+\omega^{2}(\gamma_{\bf k}/2)^2}.
\end{align}
The evolution of the spectral function $\rho_{\bf k}(t,t')$ starting from a thermal background at $T/m=2.86$ and
$\lambda=6$ is shown for two different
kernel lengths, $mt_{\rm cut}=28$ (green/grey) and $mt_{\rm cut}=84$ (black), in Fig.~\ref{fig:plotdampprop} (left). Overlaid, although barely discernible, a
fit of the form (\ref{spectrform}). As can be seen, the fit is excellent and it is the same fit for the two kernel
lengths. In Fig.~\ref{fig:plotdampprop} (right) we show the result of the approximate Wigner transforms for the
spectral function of the spatial zero mode at $m\mathcal{T}= 200$ and with two different kernel lengths. For the long
kernel case, one can nicely fit a Breit-Wigner form, 
which gives
the same damping rates and masses as in the fit of the
left-hand plot. For the short kernel case, the Breit-Wigner fit is not so accurate. 
In the following analysis, we extract the damping rates and masses from fits directly to the
time-representation of the spectral function $\rho_\mathbf{k}(t,t^\prime)$, with kernel length $mt_{\rm cut}=28$.
%
%
%
\par
Fig.~\ref{fig:plotdampall} shows the dependence on temperature of the effective mass (left) and damping rate (right)
from fits of the form 
(\ref{spectrform}) for the spectral function zero-mode
$\rho_\mathbf{0}(t,t^\prime)$,
together with the fits (\ref{evolmf}) for the mean field discussed earlier. 
For $\rho_\mathbf{0}(t,t^\prime)$, we show 
only
the results for the case of the ``basketball''
approximation,
since, for the values of the mean field considered here, it is practically zero
in the two-loop approximation.
The spectral-function zero-mode mass and damping rate
(plotted with triangles) closely follow the values for the mean field. 

\subsection{Broken phase: equilibration}
A similar analysis can be carried out in the broken phase, where there is a non-zero mean field present. In this case we
can use both the two-loop and ``basketball'' approximations to study the damping of the correlators. From the
point of view of perturbation theory, there is no damping in the two-loop approximation, for which only the perturbative ``leaf'' and the ``eye'' diagrams contribute 
to the self-energy, and their imaginary parts vanish on-shell (see also appendix B). Our task will be to study the
damping in the two-loop approximation from the $\Phi$-derived equations of motion. These formally take into account the
contributions from all orders in perturbation theory that result from any iterated insertion of the ``leaf'' and ``eye''
 diagrams into the self-energy. These diagrams contain off-shell scattering effects that can, in principle, lead to
a total non-zero on-shell imaginary part for the
self-energy, thus providing damping. For the case of the ``basketball'' approximation, the ``sunset'' diagram enters in the self-energy \eref{SEth},
which contributes to damping even in perturbation theory \cite{Jeon:1992kk,Wang:1996qf}. The solution of the
$\Phi$-derived equations of motion leads to additional contributions from higher orders compared to perturbation
theory, and thus one expects to find a larger damping and faster equilibration.
\par
Approximations based on truncations of the loop expansion of the 2PI effective action suffer from instabilities which
make it impossible to treat very large couplings and very large energy densities or particle numbers. In this sense, the
$\Phi$-derived equations of motion can be thought of as resummed perturbative, useful in the domain of weak
coupling and small fields. In the symmetric-phase simulations described previously, $\lambda=6$ is in the upper end of what stays
stable in our experience, whereas we can use temperatures (or energy densities corresponding to temperatures) up to $T/m=6$ or even
higher. In the broken phase, the instabilities turn out to be even more constraining. In particular, we will need to use
a smaller coupling ($\lambda=1$) and temperatures below $T/m=2$. As we have seen, the latter is not much of a
restriction since it still covers the region where cut-off effects are 
small.
However, it implies
that
equilibration and damping takes much longer (the damping times scale roughly as $\lambda^2 T^2$ for the sunset diagram). In particular we need to use a longer time kernel (we use $mt_{\rm cut}=84$) and we will not be able to track the evolution far enough to see chemical equilibration. We shall content ourselves with establishing kinetic equilibration and studying the damping of the mean field and the modes of the spectral function. We have no doubt that chemical equilibration will take place as well. In particular, we will see that total particle number is not conserved.
\par
The mean field is taken initially to be at the tree-level value $\phi(t=0)=v_\ts{tree}$. This is not the self-consistent finite temperature solution of the truncated equations of motion, but a bit displaced from it. Due to this initial displacement, the mean field will oscillate and damp to its equilibrium value. For the propagators we will use thermal initial conditions, as well as top-hat 1 (T1). Notice that the input mass is the broken phase zero temperature mass $\sqrt{2}m$ rather than the symmetric phase one. All results are still in units of $m$. 
\begin{figure}
\begin{center}
\epsfig{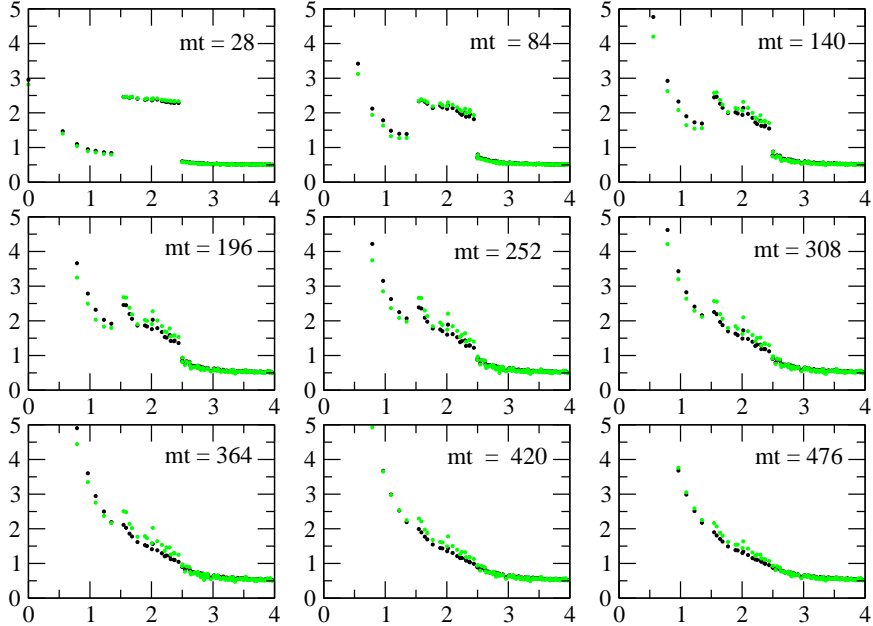}
\caption{Evolution of the occupation numbers ($n_{\bf k}$ {\it vs.} $\omega_{\bf k}/m$), starting from the T1 initial
condition, in the broken phase ($\lambda=1$) and in the two-loop (green/grey) and ``basketball'' (black) approximations, respectively.}
\label{fig:broknk}
\end{center}
\end{figure}
\begin{figure}
\begin{center}
\epsfig{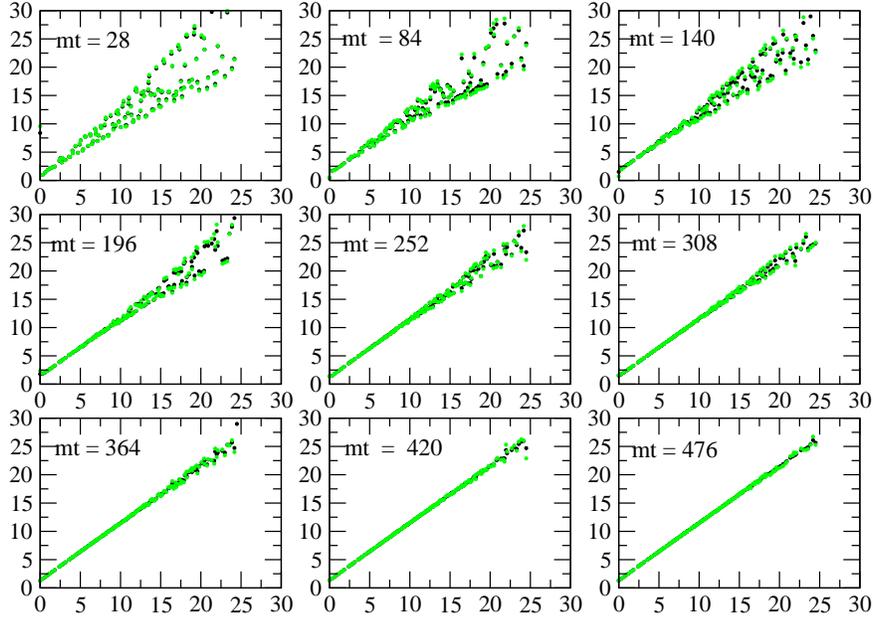}
\caption{Evolution of the dispersion relation ($\omega_{\bf k}^{2}/m^{2}$ {\it vs.} $k^{2}/m^{2}$), with the T1 initial
condition, in the two-loop (green/grey) and basketball (black) approximations, respectively.}
\label{fig:brokdisp}
\end{center}
\end{figure}
\par
\begin{figure}
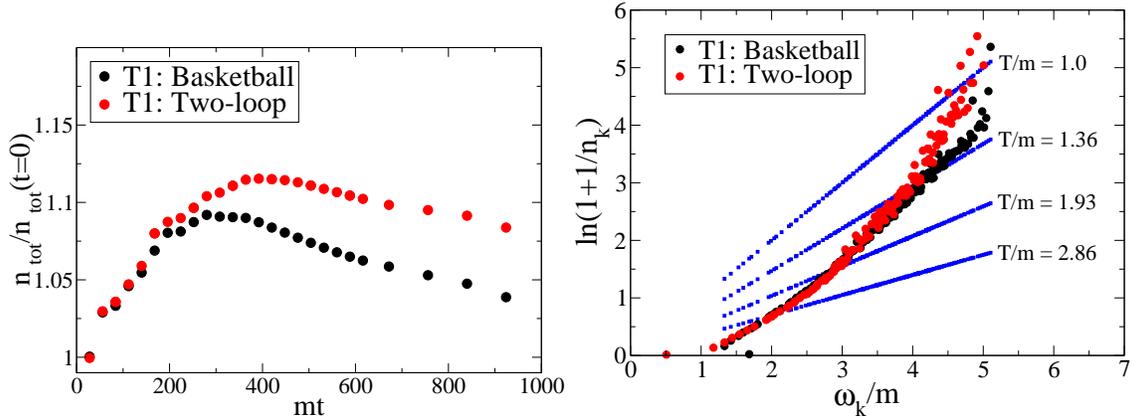

\begin{center}
\epsfig{file=pictures/P11_nktot_brok.eps,width=0.45\textwidth,clip}
\epsfig{file=pictures/P11_comp_t1000_brok_ooe.eps,width=0.45\textwidth,clip}
\caption{Left: The total particle number density $n_\ts{tot}(t)$ for the two truncations, normalized to the initial value. Right:
Particle distributions at the latest time $mt=1000$, 
for T1 and thermal initial conditions.}
\label{fig:nktotbrok}
\end{center}
\end{figure}

The evolution of the occupation numbers and the dispersion relation for
both the two-loop and ``basketball'' approximations are
shown in Fig.~\ref{fig:broknk} and Fig.~\ref{fig:brokdisp}.
Both cases show that (kinetic) equilibration is taking place. In the ``basketball'' case, equilibration is
slightly faster. 
Interestingly, the off-shell scattering effects taken
into account by the 2PI effective action with only the eye diagram lead
to an equilibration almost as fast as in the ``basketball'' case. Chemical equilibration happens on much longer time scales, and although we found that the total particle number does change in time (Fig.~\ref{fig:nktotbrok}, left), the reach of our simulations was insufficient to estimate the asymptotic temperature. At our latest time of $mt=1000$, the distribution is consistent with a Bose-Einstein with $T/m=1.24$ and $\mu/m=1.12$ (Fig.~\ref{fig:nktotbrok}, right).


\subsection{Broken phase: Mean field damping and the spectral function}
For weak coupling we expect the position of the equilibrium mean field expectation value $v$ to be close to the initial
value $\phi(0)=v_\ts{tree}$. In the case of thermal initial conditions, the initial mean field displacement corresponds to a small perturbation. As in the symmetric phase case, one can study the evolution of the mean field by linearizing the equation of motion around the equilibrium value. We write $\phi(t)=v+\sigma(t)$, where $\sigma(t)$ is the deviation. The linearized equation of motion for $\sigma$ is then
\be
\overset{..}{\sigma}(t)+M^2(T, t)\sigma(t)+\int_0^{t}dt^\prime\,\tilde{\Sigma}_{\mathbf{0}}^\rho(t,t^\prime)\sigma(t^\prime)=0.
\label{meanfeomsbro}
\ee
The vacuum expectation value $v$ is the solution of 
\be
M^2(T,t) v-\frac{\lambda}{3}v^3+\int_0^{t}dt^\prime\,\tilde{\Sigma}_\mathbf{0}^\rho(t,t^\prime)v=0.
\label{expectationvalue1}
\ee
For weak coupling
\be
v\approx \sqrt{\frac{3M^2(T)}{\lambda}}.
\label{expectationvalue2}
\ee
In Eqns.(\ref{meanfeomsbro}-\ref{expectationvalue2}), $M(T,t)$ corresponds to the finite temperature Hartree
effective mass in the broken phase, given by
\be
M^2(T,t)=-m^2+\frac{\lambda}{2}v^2+\frac{\lambda}{2}\int \frac{d^3\;k}{(2\pi)^3}F_{\mathbf{k}}(t,t)-\delta m^2.
\label{Hartreemass2}
\ee
The analysis of the
evolution of $\sigma$ proceeds as in the case of the symmetric
phase. For weak enough coupling the mean field damping rate is approximately given by the perturbative estimate
\eref{dampingrate}.
\begin{figure}
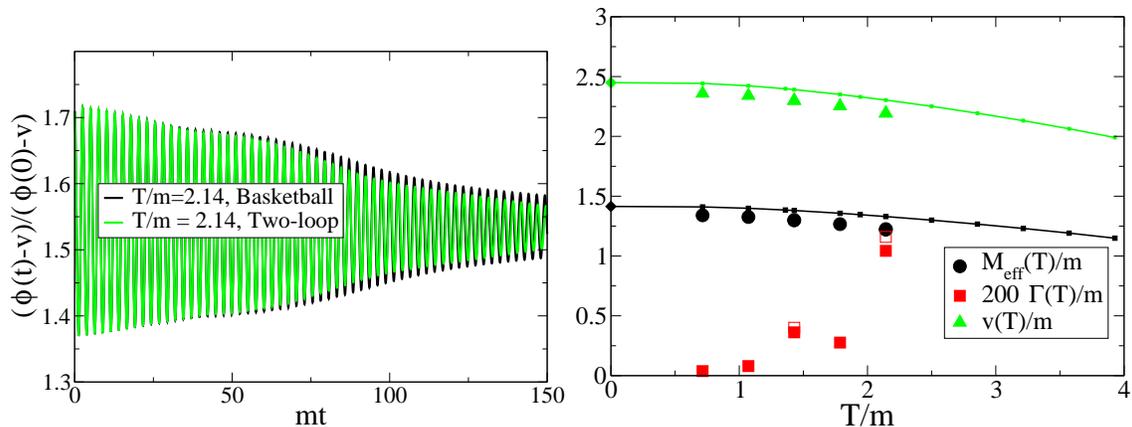

\begin{center}
\epsfig{width=0.45\textwidth,file=pictures/P12_mf_ex_brok.eps,clip}
\epsfig{width=0.45\textwidth,file=pictures/P12_mfmGv_brok.eps,clip}
\end{center}
\caption{Left: Evolution of the mean field in the broken phase in a thermal background at $T/m=2.14$, in the
two-loop and ``basketball'' approximations. Right: Effective masses, vacuum expectation values and 
damping rates
from fits of the form (\ref{evolmf}) for both 
the two-loop (empty symbols) and ``basketball'' (filled symbols)
approximations.
The lines represent the Hartree approximation.}
\label{fig:mfevolbroken}
\end{figure}
Fig.~\ref{fig:mfevolbroken} (left) shows the mean field evolution at $T/m=2.14$, in the two-loop and ``basketball''
approximations, respectively. In both cases, the mean field performs a damped oscillation, from which we can extract an
effective mass and mean field expectation value. As we can see, the damping does not follow a simple exponential form, and curiously, the two-loop data appear to indicate faster damping than the
basketball data.
Still, 
we shall use an exponential fit as a rough estimate of the damping rate.
The temperature dependence of these frequencies and damping rates, in addition to the field expectation values, is shown in Fig.~\ref{fig:mfevolbroken} (right) for
the two-loop (open symbols) and ``basketball'' (filled symbols) approximations. The masses and field expectation values are 
indistinguishable
for the two truncations. 
They
are slightly off the respective Hartree estimates 
(\ref{expectationvalue1}) and (\ref{Hartreemass2})
(full lines), indicating as in the previous section that the contribution of the ``sunset'' diagram in the
``basketball'' approximation is small relatively
to the Hartree case.
The damping rates in the two approximations are consistent with each other. 
\par
\begin{figure}
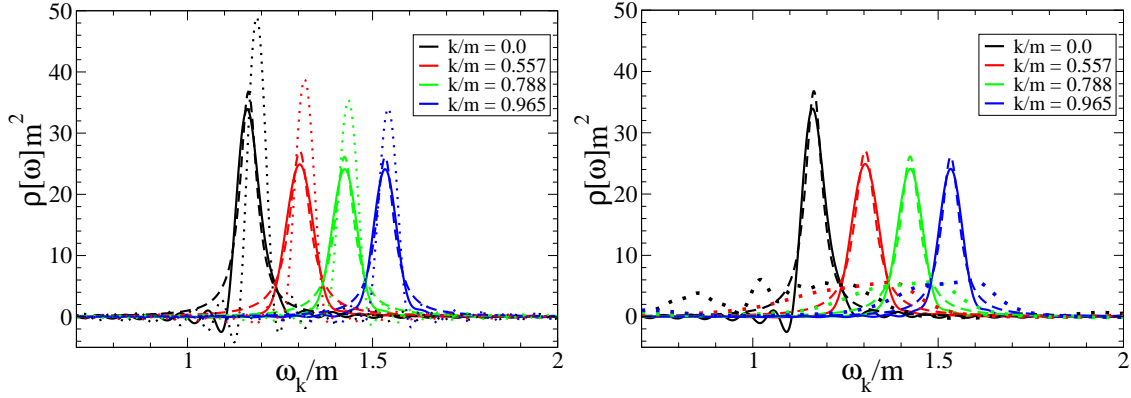

\begin{center}
\epsfig{width=0.45\textwidth,file=pictures/P12_wign_tsu_s_brok.eps,clip}
\epsfig{width=0.45\textwidth,file=pictures/P12_wign_tsu_s_brok_2.eps,clip}
\caption{Wigner transforms of the spectral function for the four lowest modes, starting from a T1 initial condition, at
two times (left plot, $m\mathcal{T}=252$ and right plot, $m\mathcal{T}=1000$) for the ``basketball'' (full lines) and
two-loop (dotted lines) approximations. The weak $\mathcal{T}$-dependence of the former shows that the system is close
to equilibrium. Indeed, Breit-Wigner fits work nicely (dashed lines). 
 The strong $\mathcal{T}$-dependence in the two-loop case suggests that the system is not yet
sufficiently close
to equilibrium.} 
\label{fig:brokwign}
\end{center}
\end{figure}

 Similarly, we observed damping in the evolution of the spectral function $\rho_{\bf k}(t,t^\prime)$ in both
 approximations.  This damping is small and not well approximated by an exponential form. 
We performed the Wigner transformation as specified in Eq.~\eref{wignertrans}, see Fig.~\ref{fig:brokwign}
to the data of
the late time evolution starting from the T1 initial condition. 
The value of the cut-off $t_\ts{cut} = 84\, m^{-1}$ produces some noise, but 
for the ``basketball'' approximation (full lines) 
there is clearly a 
well determined peak with a
finite width 
at all times,
which
can be 
fit with a Breit-Wigner form (dashed lines).
In the basketball approximation,
the form of the Wigner transform does not change much in time, which indicates that the system is relatively close to equilibrium. 
However, the results are different for 
the approximate Wigner transforms in the two-loop
approximation (dotted lines). In this case the 
transformed 
spectral function changes significantly in time from thin peaks early on
to less clear maxima at later times, still localized around the peaks, see Fig~\ref{fig:brokwign} (right). 
We do not fully understand the reason for this discrepancy;
it could be, that the cut-off time 
$t_\ts{cut} = 84\, m^{-1}$ in our implementation of the Wigner transform 
is too short in case of the two-loop approximation. 
On the other hand, the large $\mathcal{T}$-dependence of the spectral function
indicates that the system may not be sufficiently close to equilibrium
in the two-loop approximation, in which case the use of the 
approximate Wigner transform \eref{wignertrans} is questionable. 
The approximation itself
appears to work reasonably well, as shown in Figs.\ \ref{fig:nktotbrok} 
and  \ref{fig:mfevolbroken}.


\section{Conclusions}
We have studied equilibration in $\varphi^4$ theory in 3+1 dimensions for a variety of 
initial conditions, both in the symmetric and
broken phase.
Two different $\Phi$-derivable approximations including scattering effects have been used: 
two-loop and 
``basketball'', the latter corresponding to a truncation of the 2PI effective action at $\mathcal{O}(\lambda^2)$. 
In the symmetric phase the two-loop and the ``basketball'' approximations differ in that the first includes damping into
the evolution of the mean field only, 
whereas in 
the second it is also present in the equation of motion for the
2-point functions. In the broken phase both approximations include scattering effects into the 2-point functions and thus can lead to equilibration.
\par 
From the numerical study of the evolution of the occupation numbers
we were able to establish that
in the symmetric phase 
both \emph{kinetic} and \emph{chemical} equilibration is taking place, the latter at a 
substantially slower rate.
\par
By analyzing various initial conditions we found that, after kinetic
equilibration, the occupation numbers at intermediate times are
given by a Bose-Einstein distribution with an effective chemical potential.  
This is
similar to what was found in previous studies in 2+1 dimensions
\cite{Juchem:2003bi}. Given the same total energy, we found the intermediate effective chemical potential to be generally
non-zero, with its size related to the initial total particle numbers, as may be expected.
Eventually, one would expect the limit
distribution to depend only on the energy density of the system, and
close to a Bose-Einstein with zero chemical potential \cite{Berges:2000ur,Juchem:2003bi}. Comparing to the studies
in 2+1 dimensions \cite{Juchem:2003bi}, our numerical analysis indicates that the subsequent
chemical equilibration is much slower in 3+1 dimensions.
\par 
We were also able to extract 
effective masses and damping rates
from the analysis of the evolution of the
mean field and the spectral function. The 
contributions to the mass from the two-loop and the ``basketball''
approximations seem to be small comparing to the Hartree case. In the symmetric phase, the results for the damping rate are about a 20-40$\%$
higher than the perturbative estimates. This 
indicates
that the scattering effects associated with the
resummations
encoded in the $\Phi$-derivable approximation are substantial. Finally, we checked that the damping rate obtained from the mean
field coincides with the one from the spectral function zero-mode. 
\par
In the broken phase we found that both the two-loop and the ``basketball'' approximation lead to equilibration.
Surprisingly, the equilibration seems to be just a bit slower in the two-loop case. This is particularly remarkable,
since, in perturbation theory, the two-loop approximation does not have on-shell damping. 
Indeed, in that case only the perturbative ``leaf''
and ``eye'' self-energy diagrams contribute, and their imaginary parts vanish on-shell (see appendix B). The fact that
the two-loop approximation in $\varphi^4$ theory equilibrates so fast might be relevant to pure gauge theories, where
the lowest order $\Phi$-derivable approximation (at $\mathcal{O}(g^2)$) considers the same diagrams. 
\par 
On a practical note, we found that the loop expansion suffers from restrictions reminiscent of perturbative expansions,
in that large couplings and/or large field occupation numbers trigger instabilities when solving the equations of
motion. This has  
to our knowledge
not been reported for simulations in 1+1 and 2+1 dimensions,
although we have found them in those cases
as well. 
In addition to the instabilities, issues such as
CPU time and computer memory necessary for dealing with the memory integrals are significant restrictions, especially
when studying late time thermalization. Expansions in $1/N$ with $N$ the number of fields have been shown to be more
stable and able to deal with non-perturbatively large occupation numbers at large coupling
\cite{Berges:2001fi,Berges:2002cz,Mihaila:2003mh,Arrizabalaga:2004iw}. In such cases care should be taken to ensure that the expansion is controlled by using a sufficiently large value of $N$ \cite{Aarts:2001yn}.


\begin{acknowledgments} 
We would like to thank Gert Aarts, J\"{u}rgen Berges, Szabolcs Bors\'{a}nyi,
Julien Serreau and Urko Reinosa for useful discussions. 
We thank Fokke Dijkstra, SARA and NCF for help in paralellizing the 
computer code.
A.T. is supported by PPARC SPG {\it``Classical lattice field theory''}. Part of this work was conducted on the SGI
Origin platform using COSMOS Consortium facilities, funded by HEFCE, PPARC and SGI. This work also received support from 
FOM/NWO.
\end{acknowledgments}
\appendix
\section*{Appendix A: Energy-momentum tensor in $\Phi$-derivable approximations}

The energy-momentum tensor for a given truncation of the 2PI effective action
can be determined through Noether's procedure, i.e.~by identifying the current
term resulting
from the space-time dependent translations $x^\mu\rightarrow
x^\mu+\epsilon^\mu(x)$. A convenient way to find the Noether current is to
pertorm an infinitesimal translation that vanishes on the space-time boundary. 
The translation $x^\mu\rightarrow
x^\mu+\epsilon^\mu(x)$ can be viewed as a transformation of the relevant variables, which in the case
of the 2PI effective action are the mean field $\phi(x)$ and the connected
2-point function
$G(x,y)$. This transformation is given by
\begin{align}
\phi(x)&\longrightarrow \phi^\prime(x)\equiv \phi(x+\epsilon(x))=\phi(x)+ \epsilon^\lambda(x)\partial^x_\lambda \phi(x), 
\label{meanfieldtrafo} \\
G(x,y)&\longrightarrow G^\prime(x,y)\equiv G(x+\epsilon(x),y+\epsilon(y))=G(x,y)+\epsilon^\lambda(x)\partial^x_\lambda
G(x,y)+\epsilon^\lambda(y)\partial^y_\lambda
G(x,y) \label{proptrafo},
\end{align}
where the variables that the partial derivatives act on are indicated with a
superscript. Under these transformations the variation of the 2PI effective action
$\Gamma[\phi,G]$ can be formally written as
\be
\delta \Gamma [\phi,G]=\int_x\; T^{\mu\nu}(x) \partial_\mu \epsilon_\nu(x),
\label{formaldefemt}
\ee
where $\int_x\equiv\int d^4x$. The quantity $T^{\mu\nu}$ defines a conserved
 Noether current, which is identified as the energy-momentum tensor. 
To see that it is indeed conserved, notice that when the $\Phi$-derived
equations of motion for $\phi$ and $G$ are satisfied.
This applies, in particular, to the transformations
\eref{meanfieldtrafo} and \eref{proptrafo}, and hence $\delta
\Gamma[\phi,G]=0$. Taking \eref{formaldefemt} and making a partial
integration one obtains
\be
\delta \Gamma[\phi,G]=-\int_x\;\epsilon_\nu(x)\,\partial_\mu T^{\mu\nu}(x)=0.
\ee
Since $\epsilon_\nu(x)$ can be taken arbitrary, the energy-momentum tensor is
conserved, i.e. $\partial_\mu T^{\mu\nu}(x)=0$.\\
Below we give the explicit form of the energy-momentum tensor $T^{\mu\nu}$ for any
$\Phi$-derivable approximation by applying the transformations
(\ref{meanfieldtrafo}-\ref{proptrafo}) and using the definition
\eref{formaldefemt}.
We study independently the contributions
coming from the four terms in the action \eref{2PIfunctional}:
\begin{itemize}
\item[i)] The first term in Eq.~\eref{2PIfunctional} is given by $S[\phi]$ and leads to the usual form of the energy-momentum tensor for
the mean field $\phi$, namely
\be
T_1^{\mu\nu}(x)=\partial^\mu\phi(x)\partial^\nu\phi(x) - g^{\mu\nu}\left[\half
\partial_\lambda\phi(x)\partial^\lambda\phi(x)-\half
m^2\phi(x)^2-\frac{\lambda}{4!}\phi(x)^4\right].
\ee
\item[ii)] The second term appearing in Eq.~\eref{2PIfunctional} does not contribute to
the energy-momentum tensor. Indeed, applying the transformations \eref{proptrafo}
leads to
\begin{multline}
\delta\left[ {\rm Tr}\ln G\right]
={\rm Tr}G^{-1}\delta G\\
=\int_x\int_y
G^{-1}(y,x)\left[\epsilon^\lambda(x)\partial_\lambda G(x,y) +
(x\leftrightarrow y) \right]
=2\int_x \epsilon^\lambda(x)\partial_\lambda^x
\delta^{(4)}(x-x)=0.
\end{multline}
Hence, the corresponding contribution $T^{\mu\nu}_2(x)$ to the energy-momentum tensor vanishes.
\item[iii)] To obtain the contribution from the third term, given by $(i/2)\,\mbox{Tr}
\left[G_0^{-1}-G^{-1}\right]G$, one proceeds similarly to what was done in i).
Under the transformation \eref{proptrafo}, it becomes
\begin{multline}
\delta \left[\frac{i}{2}\mbox{Tr}
\left[G_0^{-1}-G^{-1}\right]G \right]=\\
\frac{1}{2}\int_x\int_y
\left(-\partial_\mu^x\partial^\mu_x-m^2-\frac{\lambda}{2}\phi(x)^2\right)\delta^{(4)}(x-y)\left[\epsilon^\lambda(x)\partial_\lambda^xG(x,y)+(x\leftrightarrow y)\right]\\
-\frac{1\lambda}{2}\int_x\int_y
\phi(x)\epsilon^\lambda(x) \partial_\lambda^x\phi(x) \delta^{(4)}(x-y)G(x,y).
\end{multline}
After some partial integrations and making use of the identity 
\be
\int_y \partial_x^\mu\left[ \delta^{(4)}(x-y)G(x,y)\right]=\int_y
\delta^{(4)}(x-y)\left[\partial_x^\mu G(x,y)+\partial_x^\mu G(y,x)
\right],
\ee
one can write the above in the form given by Eq.~\eref{formaldefemt}, which allows to extract
the contribution of this term to the
energy-momentum tensor. One finds
\be
T_3^{\mu\nu}(x)=\int_y \delta^{(4)}(x-y)\left[\partial^\mu_x\partial^\nu_y-\half
g^{\mu\nu}\partial^\lambda_x\partial_\lambda^y+\half
g^{\mu\nu}m^2+\frac{1}{4}g^{\mu\nu}\lambda\phi(x)^2\right]
G(x,y). 
\ee
\item[iv)] For the fourth term in Eq.~\ref{2PIfunctional}, which is given by the functional $\Phi[\phi,G]$, the
	transformations (\ref{meanfieldtrafo}-\ref{proptrafo}) give
	\be
\delta \Phi[\phi,G]=\int_x \frac{\delta \Phi}{\delta
\phi(x)}\epsilon_\mu(x)\partial_x^\mu \phi(x)+\int_x\int_y\frac{\delta \Phi}{\delta
G(x,y)}\left[\epsilon_\mu(x)\partial^{\mu}_x
G(y,x)+\epsilon_\mu(y)\partial^{\mu}_y G(y,x)\right].
\ee
What we want is to write this in a form similar to \eref{formaldefemt} such that its
contribution to the energy-momentum tensor can be extracted. To do this,
notice that the functional $\Phi[\phi,G]$ is a scalar quantity that does
not contain derivative terms. This means that, under the space-time
translation $x^\mu\rightarrow
x^\mu+\epsilon^\mu(x)$, the terms in $\Phi$ only change by the appearance of the Jacobian of the
transformation at every loop integration. This Jacobian can be accommodated by
a simultaneous change in a scale factor $\zeta(x)$ introduced at every
integration vertex as $\lambda \rightarrow \zeta(x)\lambda$ \cite{Baym:1962sx,Ivanov:1998nv}. Thus the
simultaneous variation 
\begin{equation}
\phi(x)\rightarrow \phi(x+\epsilon(x)),\quad G(x,y)\rightarrow
G(x+\epsilon(x),y+\epsilon(y)),\quad \zeta(x)=1\rightarrow
\zeta(x)=\mbox{det}\left(\delta^{\mu}_\nu+\partial_\nu^x \epsilon^\mu(x)\right) 
\end{equation}
leaves the functional $\Phi$ invariant. For infinitesimal transformations,
this invariance implies
\be
\int_x \frac{\delta \Phi}{\delta
\phi(x)}\epsilon_\mu(x)\partial_x^\mu \phi(x)+\int_x\int_y\frac{\delta \Phi}{\delta
G(x,y)}\left[\epsilon_\mu(x)\partial^{\mu}_x
G(y,x)+\epsilon_\mu(y)\partial^{\mu}_y G(y,x)\right]+\int_x
\frac{\delta\Phi}{\delta \zeta(x)} \partial_\mu^x \epsilon^\mu(x)=0.
\label{generalizedtrafoinphi}
\ee
One can then use the identity \eref{generalizedtrafoinphi} to write
the variation $\delta\Phi[\phi,G]$ in a form similar to \eref{formaldefemt} as
\be
\delta \Phi[\phi,G]=-\int_x
\frac{\delta\Phi}{\delta \zeta(x)}\Big|_{\zeta=1} \partial_\mu^x
\epsilon^\mu(x).
\ee
In this manner, the contribution of the functional $\Phi$ to the energy
momentum tensor can be written as
\be
T^{\mu\nu}_4(x)=-g^{\mu\nu}\frac{\delta\Phi}{\delta\zeta(x)}\Big|_{\zeta=1}.
\ee

\end{itemize}
The total energy-momentum tensor is obtained by adding up all
the contributions from i)-iv),
i.e.~$T^{\mu\nu}(x)=T_1^{\mu\nu}(x)+T_2^{\mu\nu}(x)+T_3^{\mu\nu}(x)+T_4^{\mu\nu}(x)$.
The result can be compactly written as
\begin{multline}
T^{\mu\nu}(x)=\left[\partial_x^\mu\partial^\nu_y-\half
g^{\mu\nu}\partial_\lambda^x\partial_y^\lambda+\half
g^{\mu\nu}m^2\right]\big(\phi(x)\phi(y)+G(x,y)\big)\Big|_{x=y}\\
+g^{\mu\nu}\frac{1}{4!}\lambda
\phi(x)^4+\frac{1}{4}g^{\mu\nu}\phi(x)^2
G(x,x)-g^{\mu\nu}\frac{\delta\Phi}{\delta \zeta(x)}\Big|_{\zeta=1}.
\end{multline}


\section*{Appendix B: Perturbative damping from the "eye" diagram}

For completeness, we include here the calculation of the imaginary part of the perturbative ``eye'' diagram in equilibrium in the real-time formalism, using
the Schwinger-Keldysh contour. 
More 
details can be found in e.g.\ \cite{LeBellac}. We introduce the labels $+$ or $-$ to indicate whether the time variables of any
quantity live respectively on
the $\mathcal{C}^+$ or $\mathcal{C}^-$ branch of the contour. In terms of the various contour components, the
retarded self-energy is given by $\Sigma^R(x,y)=\Sigma^{++}(x,y)+\Sigma^{+-}(x,y)$. For the case of the eye-diagram, in
momentum space one has
\be
\Sigmaeye^R(p)=\Sigmaeye^{++}(p)+\Sigmaeye^{+-}(p)=\frac{i\lambda^2 v(T)^2}{2}\int_k
\big[G^{++}(k)G^{++}(k-p)-G^{+-}(k)G^{-+}(k-p) \big],
\label{eyepert}
\ee
where $v(T)$ is the mean field equilibrium expectation value at temperature $T$. We shall use $\int_k$ and
$\int_\mathbf{k}$ to denote the 4- and 3-dimensional momentum integrations $\int d^4 k/(2\pi)^4$ and $\int d^3
k/(2\pi)^3$ respectively.
\par 
It is convenient to use the Keldysh basis \cite{Keldysh:1964,Chou:1985}, where the various components of $G$ are given in terms of
the symmetric, retarded and advanced correlators $F$, $G^R$ and $G^A$ respectively. Their perturbative expressions are 
\begin{align}
F(k)&=2\pi\delta(k^2-m^2)\left[n(k_0)+\half\right], \label{Fpert}\\
G^R(k)&=\frac{1}{(k_0+i\epsilon)^2-\mathbf{k}^2-m^2}, \label{grpert}\\
G^A(k)&=\frac{1}{(k_0-i\epsilon)^2-\mathbf{k}^2-m^2},\label{gapert}
\end{align}
with $\epsilon=0^+$ and $n(k_0)$ the Bose-Einstein distribution at temperature $T$ and energy $k_0$.
\par
In the Keldysh basis the retarded self-energy \eref{eyepert} becomes
\be
\Sigmaeye^{R}(p)=\frac{i\lambda^2v(T)^2}{2}\int_k \Big[ \half
\left( G_R(k)G_R(k-p)+G_A(k)G_A(k-p) \right) 
-iF(k)G_R(k-p)-iG_A(k)F(k-p)\Big].
\ee
The first two terms in the RHS have poles at only one side of the complex
plane. In the integration over $k_0$ one can always choose to close the
contour at the other side, thus these two terms vanish. The last two
contributions can be seen to be equal to each other by the change of
variable $\mathbf{k}\leftrightarrow (\mathbf{p}-\mathbf{k})$. After performing the $k_0$ integration with the help
of the $\delta$-functions in $F$ we obtain
\be
\begin{split}
\Sigmaeye^{R}(\omega,\mathbf{p})=\lambda^2v(T)^2\int_{\mathbf{k}}\frac{1}{2\omega_{k}}\left(
n_{k}+\half\right) \Bigg[
\frac{1}{(\omega-\omega_\mathbf{k})^2-\omega_{\mathbf{p-k}}^2-i\epsilon\,\mbox{sgn}(\omega-\omega_{k})}\\
+\frac{1}{(\omega+\omega_\mathbf{k})^2-\omega_\mathbf{p-k}^2-i\epsilon\,\mbox{sgn}(\omega+\omega_\mathbf{k})}\Bigg].
\label{k0integration}
\end{split}
\ee
The imaginary part of the self-energy is obtained by using $1/(x+i\epsilon)=P(1/x)-i\pi\delta(x)$ and decomposing the
resulting delta functions. For $\omega >0$, and after convenient changes of variable, we obtain
\be
\mbox{Im}\Sigmaeye^{R}(\omega,p)=\frac{\lambda^2 v(T)^2}{2}\int_{\mathbf{k}}\frac{\pi}{4\omega_\mathbf{k}\omega_\mathbf{p-k}}\left\{
\left[2n_\mathbf{k}+1\right]\delta(\omega-\omega_\mathbf{k}-\omega_\mathbf{p-k})+2n_\mathbf{k}\delta(\omega+\omega_\mathbf{k}-\omega_\mathbf{p-k})
\right\}.
\label{imeyeintegral}
\ee
The first contribution to the integral corresponds to the \emph{decay} of an off-shell excitation into
two on-shell
excitations. The second one corresponds to \emph{Landau damping} via scattering of the off-shell excitation with
on-shell particles from the heat bath (occuring only at $T\neq 0$).
\par 
One can make use of the delta functions present
in (\ref{imeyeintegral}) to solve the angular part of the integral over the
internal momentum $\mathbf{k}$. Indeed, using the property 
\be
\delta\big(f(x)\big)=\sum_{\ts{roots}}
\frac{\delta(x-x_{\ts{root}})}{|f^{\prime}(x_{\ts{root}})|},
\label{deltaf}
\ee
one can solve the angular part of the integral if $f(x)$ is taken to be
$\omega\pm \omega_\mathbf{k}-\omega_{\mathbf{p}-\mathbf{k}}$ with $x=\cos \theta$ and $\theta$ the angle between the
vectors $\mathbf{k}$ and $\mathbf{p}$ (the $+$ sign corresponds to decay and the $-$ sign to Landau damping). The sum
present in (\ref{deltaf}) is over the roots of the function $f(x)$, which for
$f(x)=\omega\pm \omega_\mathbf{k}-\omega_{\mathbf{p}-\mathbf{k}}$, are given by
\be
x=\frac{\mathbf{p}^2-\omega^2\mp2\omega\,\omega_\mathbf{k}}{2|\mathbf{p}||\mathbf{k}|},
\ee
with $|f^{\prime}(x)|=|\mathbf{p}||\mathbf{k}|/\omega_{\mathbf{p}-\mathbf{k}}$. In order to obtain a
nonzero contribution, the roots of the function $f(x)$ in the $\delta$'s must
be inside the interval $[-1,1]$, i.e.~
\be
-1 \le
\frac{\mathbf{p}^2-\omega^2\pm2\omega\,\omega_\mathbf{k}}{2|\mathbf{p}||\mathbf{k}|} \le
1,
\label{angularrestriction}
\ee
We analyze the two regions independently:
\begin{enumerate}
\item \emph{Decay:} In the case of decay the $\delta$-function in
Eq.~(\ref{imeyeintegral}) implies that
$\omega=\omega_\mathbf{k}+\omega_{\mathbf{p}-\mathbf{k}}$. 
This is only possible provided $\omega \ge \sqrt{\smash[b]{\mathbf{p}^2+4m^2}}$,
so this contribution to damping only occurs
above the 2-particle threshold. The lower and upper integration limits $k_-$
and $k_+$, which result from the restriction
(\ref{angularrestriction}), can be easily expressed as
\be
k_{\pm}=\left|
\pm\frac{|\mathbf{p}|}{2}+\frac{\omega}{2}\sqrt{1+\frac{4m^2}{\mathbf{p}^2-\omega^2}}\right|.
\label{integrationlimits}
\ee
\item \emph{Landau damping:} In this case the contribution only
occurs below the light-cone ($\mathbf{p}^2>\omega^2$). The integration limits resulting from the
restriction (\ref{angularrestriction}) turn out to be the same as in the case
of decay\footnote{This is 
not true in general. It does not happen, for
instance, in the contribution of the sunset diagram to damping \cite{Wang:1996qf,Jeon:1995if}.}. Notice that both for decay and Landau damping the function
inside the square root in the integration
limits (\ref{integrationlimits}) is positive, so $k_{\pm}$ are real. 
\end{enumerate}
After the angular integration is performed, the contributions to the imaginary part of the retarded self-energy coming
from decay and Landau damping can thus be written as
\be
\mbox{Im}\Sigmaeye^{R}(\omega,p)=\frac{\lambda^2v(T)^2}{4}\int^{k_+}_{k_-}\frac{dk}{4\pi}
\frac{k}{|\mathbf{p}|}\frac{1}{2\omega_\mathbf{k}}\left\{\left[2n_\mathbf{k}+1\right] \Theta(\omega^2-\mathbf{p}^2-4m^2)+
 2n_\mathbf{k}\Theta(\mathbf{p}^2-\omega^2)\right\}
 \ee
The remaining integrations can be easily performed to obtain
\begin{multline}
\mbox{Im}\Sigmaeye^{R}(\omega,p)=\frac{\lambda^2 v(T)^2}{16\pi|\mathbf{p}|}\Bigg\{ \left[ T\ln\left(
\frac{1-e^{-\beta\omega_+}}{1-e^{-\beta\omega_-}}\right)+\frac{1}{2}(\omega_+-\omega_-)\right]
\theta(\omega^2-\mathbf{p}^2-4m^2)+\\ 
+  T\ln\left(
\frac{1-e^{-\beta\omega_+}}{1-e^{-\beta\omega_-}}\right)\theta(\mathbf{p}^2-\omega^2)\Bigg\},
\label{im1}
\end{multline}
with $\omega_{\pm}$ given by
\be
\omega_{\pm}=\frac{1}{2}\Big|\omega\pm
|\mathbf{p}|\sqrt{1+\frac{4m^2}{\mathbf{p}^2-\omega^2}}\Big|.
\ee
The same result was obtained using Laplace transform methods \cite{Boyanovsky:1996zy}. We observe from \eref{im1} that the perturbative retarded self-energy coming
from the ``eye'' diagram does not contribute to on-shell damping. The corresponding on-shell plasma excitations (plasmons) are stable and behave as free
quasiparticles.
\par
The same conclusion can be obtained by performing the analysis of the damping rate on the lattice. In this case, an
explicit form for the damping rate such as \eref{im1} cannot be given due to, among other things, the lack of rotational
invariance. The lattice damping rate can be 
calculated by studying the evolution of 
the mean field, as done 
in section \ref{sec:symmetricphase}.


\bibliography{equilibration}
\end{document}